\documentclass{JHEP3}
\usepackage{amsmath,amsfonts,amsthm,graphicx,ulem}
\newcommand{\bal}{\begin{align}}
\newcommand{\eal}{\end{align}}
\newcommand{\beq}{\begin{equation}}
\newcommand{\eeq}{\end{equation}}
\newcommand\beqa{\begin{eqnarray}}
\newcommand\eeqa{\end{eqnarray}}
\newcommand\bea{\begin{array}}
\newcommand\eea{\end{array}}

\newcommand{\eq}[1]{(\ref{#1})}

\renewcommand{\sl}{\mathfrak{sl}}
\newcommand{\gl}{\mathfrak{gl}}

\newcommand{\Qs}{{\mathsf Q}}

\newcommand{\ps}{{\mathsf p}}
\newcommand{\Ts}{{\mathsf T}}

\newcommand{\mx}{x^{\rm mir}}

\newcommand{\ph}{{\rm ph}}
\newcommand{\tU}{ V}
\newcommand{\jA}{{\hat 1}}
\newcommand{\jB}{{\hat 2}}
\newcommand{\jC}{{\hat 3}}
\newcommand{\jD}{{\hat 4}}
\newcommand{\jE}{{\hat 5}}
\newcommand{\jI}{{\hat \imath}}
\newcommand{\jJ}{{\hat \jmath}}
\newcommand{\jF}{{\hat f}}
\newcommand{\ja}{{{1}}}
\newcommand{\jb}{{{2}}}
\newcommand{\jc}{{{3}}}
\newcommand{\jd}{{{4}}}
\newcommand{\je}{{{5}}}

\renewcommand{\leq}{\leqslant}
\renewcommand{\geq}{\geqslant}
\renewcommand{\le}{\leqslant}
\renewcommand{\ge}{\geqslant}

\newcommand{\mir}{{\rm mir}}
\newcommand{\rb}{\right)}
\newcommand{\lb}{\left(}
    \newcommand{\nn}{\nonumber}
    
    \newcommand{\COMMENT}[1]{}
    
    \newcommand{\neqa}{\nonumber\end{eqnarray}}
    \newcommand{\la}[1]{\label{#1}}
\def\eps{{\epsilon}}

\def\[{\left[}
\def\]{\right]}

\def\l{\lambda}
\def\e{\exp \epsilon}

\def\s{\sigma}

\def\b{\beta}

\def\<{\langle}
\def\>{\rangle}

\def\i2{\frac{i}{2}}

\def\Uh{{U}}

\def\p{\partial}

\newcommand{\DbO}{\mathcal{B}_1=\{\Qs_{\ja},\linebreak[1] \Qs_{\jb},\linebreak[1] \Qs_{\jc},\linebreak[1] \Qs_{\jd},\linebreak[1] q_{\jA},\linebreak[1] q_{\jB},\linebreak[1] q_{\jC}, q_\jD\}}

\newcommand{\pbar}[1]{{\overset{\multimap}{#1}}{}}
\newcommand{\mbar}[1]{{\overline{#1}^{}}}

\title{ Wronskian Solution for AdS/CFT Y-system
}
\author{ Nikolay Gromov\\King's College, London Department of Mathematics WC2R 2LS, UK \&\\ St.Petersburg INP, St.Petersburg, Russia \\
E-mail: \email{nikgromov$\bullet$gmail.com}}
\author{Vladimir Kazakov\footnote{member of Institut Universitaire de France}\\Ecole Normale Superieure, LPT,  75231 Paris CEDEX-5, France \&   \\
l'Universit\'e Paris-VI, Paris, France;\\
E-mail: \email{kazakov$\bullet$lpt.ens.fr}}
\author{Sebastien Leurent\\Ecole Normale Superieure, LPT,  75231 Paris
  CEDEX-5, France  \\
E-mail: \email{leurent$\bullet$lpt.ens.fr}}
\author{Zengo Tsuboi
\footnote{in research partnership with Okayama Institute for Quantum physics; 
present address: 
Max-Planck-Institut f\"{u}r Gravitationsphysik, 
Albert-Einstein-Institut, 
Am M\"{u}hlenberg 1, 
14476 Potsdam,  
Germany}
\\
Osaka City University Advanced Mathematical Institute,
3-3-138 Sugimoto, Sumiyoshi-ku
Osaka 558-8585 Japan  \&   \\
Okayama Institute for Quantum Physics, Kyoyama 1-9-1, Okayama 700-0015, Japan
\\
E-mail: \email{ztsuboi$\bullet$gmail.com}}
\abstract{
Using the  discrete Hirota integrability we  find the general solution of
the full quantum Y-system for the spectrum of anomalous dimensions of operators in the planar AdS\(_5\)/CFT\(_4\) correspondence in terms of Wronskian-like determinants parameterized by a finite number of Baxter's Q-functions. We consider it as a useful  step towards the  construction of a {\it finite} system of non-linear integral equations (FiNLIE) for the full spectrum. The explicit   asymptotic    form of all the  Q-functions  for the large size operators is presented. We  establish   the symmetries and the
analyticity properties of the asymptotic Q-functions and  discuss their possible generalization to any finite size operators.}
\preprint{LPT ENS-10/38}
\usepackage{varioref}
\usepackage{makeidx}
\makeindex
\usepackage{amssymb}

\begin{document}

  \section{Introduction}

The  Y-system for the full spectrum of energies/dimensions in the planar AdS\(_5\)/CFT\(_4\)  system conjectured in \cite{Gromov:2009tv}
has passed  a few important tests. It was re-derived and better understood within    the TBA approach   \cite{Gromov:2009bc,Bombardelli:2009ns,Arutyunov:2009ur} and successfully tested in the weak coupling
by   comparison with the  perturbative expansion in N=4 SYM theory up to 4-5 loops \cite{Bajnok:2008bm,Fiamberti:2008sm,Velizhanin:2008pc,Minahan:2009wg,Arutyunov:2010gb}.
Remarkably, the very same Y-system was shown to be responsible for the spectrum of
\(\beta\)-deformed \(N=1\) case in \cite{Gromov:2010dy}
where the perturbative results \cite{Fiamberti:2008sm,Fiamberti:2010fw} were reproduced up to \(11\) loops .
At strong coupling the Y-system perfectly reproduces the results
of quasi-classical quantization of  highly excited states of the superstring
in \(\sl_2\) sector, in the regime where there is no way to ignore the finite size effects   \cite{Gromov:2009tq},
 demonstrating that the formidable wrapping problem  finds its successful resolution   within the Y-system.
Furthermore, some of these results were extended to the generic finite-gap string state in \cite{Gromov:2010vb}
and to the ABJM model in \cite{Gromov:2009at}.

\begin{figure}
\begin{center}
\includegraphics[scale=0.7]{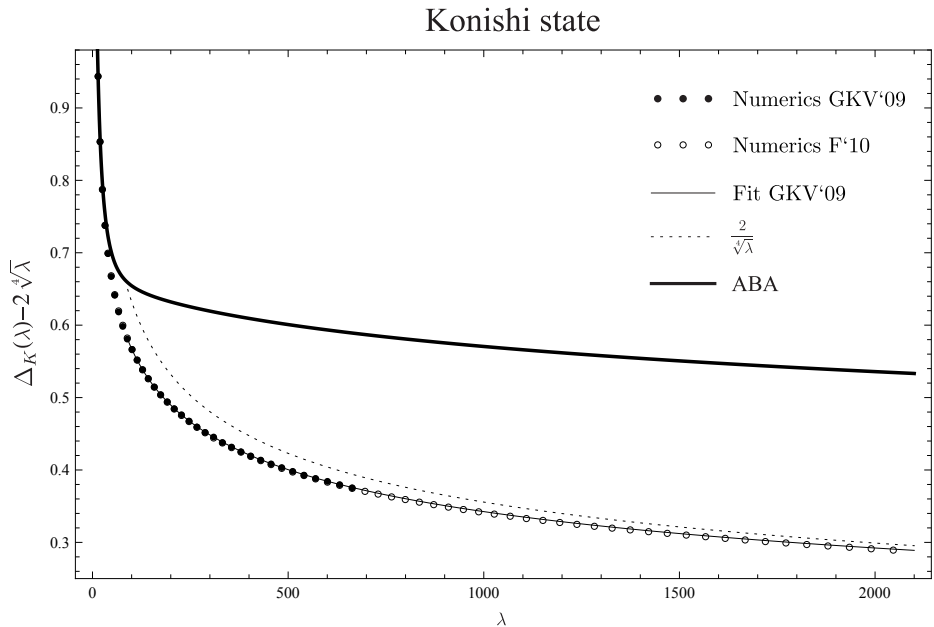}
\end{center}
\caption{\la{fig:KonishiNew}Numerical results of solution of the AdS/CFT Y-system (in integral, TBA form) for Konishi dimension as a function of `t Hooft coupling \(\l\)  (from \cite{Gromov:2009zb}  and \cite{Frolov:2010wt}). }
\end{figure}

 The
numerical calculation of the Konishi dimension from the
Y-system, combined  with the TBA approach in \cite{Gromov:2009zb},
has provided the data covering a range of values of the `t Hooft coupling
enough to confirm the leading strong coupling asymptotics \(2\l^{1/4}\)  obtained on the string side of the duality \cite{Gubser:1998bc}
and to predict, with a reasonable accuracy, the next, subleading correction as being  \(1.99\cdot\l^{-1/4}\).\footnote{ It is tempting
to think that  it is exactly \(2\cdot\l^{-1/4}\).
We hope that this coefficient  will be eventually compared with the direct worldsheet 2-loop calculations.}
A more extensive numerical study 
recently done in \cite{Frolov:2010wt} allowed to
extend the range of the interval of `t Hooft couplings in the strong coupling regime by more than \(3\) times.
This new data from  \cite{Frolov:2010wt}
perfectly confirms old results:
the new points follow   the fitting curve from \cite{Gromov:2009zb},
within the accuracy margins\footnote{This means that the fitting function presented in \cite{Gromov:2009zb}
describes all available data concerning Konishi anomalous dimension, including \cite{Frolov:2010wt},
with the accuracy \(\sim 0.001\), which was the target accuracy for \cite{Gromov:2009zb}.
At the same time,  the claimed accuracy of the recent numerics is higher.
} (see Fig.\ref{fig:KonishiNew}).

By now it is fair to say that the Y-system is the correct framework to study the
spectrum of this important duality. However there are still many open problems.
The main problem which
has  been preventing us from applying the Y-system to more complicated states
is  to convert it into a system of integral equations
suitable for numerical/analytical study at intermediate couplings.
This problem was only solved for some simple states like the Konishi state.
It is
important to understand the properties of a   general solution of Y-system
describing the physical states of AdS/CFT correspondence.

The AdS/CFT Y-system is an infinite set of functional equations on the functions of a spectral parameter,
related  to the Hirota  bi-linear difference equation
(T-system) well known in quantum integrability. It differs from the Y-systems for the previously known quantum integrable models (1D spin chains, sigma-models etc.) by the specific,  so called  \(\mathbb{T}\)-hook boundary conditions in the representation space (w.r.t. the discrete indices living on a 2D lattice \((a,s)\in \mathbb{Z}\times \mathbb{Z} \))  and complicated analyticity properties w.r.t. the spectral parameter.

On the other hand, it is known from our experience with some relativistic sigma models and spin chains that the same quantum systems  described at a finite volume by a Y-system may obey a rather different,
finite set of non-linear integral equations (which we will abbreviate here as
FiNLIE). The first example of such equations, called DdV equations, were given  by Destri and deVega for the Thirring model \cite{Destri:1987ug} followed by a considerable activity in this direction.
A new approach for the search of such FiNLIE for the integrable sigma-models  based on the integrable Hirota dynamics of the Y-system, together with a few simple assumptions on its analyticity structure, was proposed in \cite{Gromov:2008gj} and developed in \cite{Kazakov:2010kf}. It relies on the fact that the Y-system, or the underlying T-system, can be solved in terms of a finite number of \(\Qs\)-functions, analogs of those introduced by Baxter for the XXZ chain.  The T-functions can be represented in terms of finite determinants
(Wronskians\footnote{
To be precise, they are variations of a discrete analogue of Wronskian called ``Casoratian''.
}) of those \(\Qs\)-functions and, knowing the analyticity properties of \(\Qs\)'s, one can then write a FiNLIE solving the finite size problem.

Such a FiNLIE would be a very welcome progress for the study of the spectrum of   AdS\(_5\)/CFT\(_4\) system. It would provide us not only with a more efficient analytic and numerical tool for the study of this complex model but also most probably give some insight into the structure of this duality. However, due to the complexity of the \(\mathbb{T}\)-hook boundary conditions and  of the analyticity properties of Y-functions, this FiNLIE remains unknown. We propose in this paper an important, in our opinion, step towards the derivation of such equations by writing a general Wronskian-type solution of the corresponding T-system  for the \(\mathbb{T}\)-hook boundary conditions. This solution will be a natural generalization of our explicit general solution,
found by three of the current authors \cite{Gromov:2010vb} for  a simplified  T-system\footnote{called in mathematical literature the Q-system, with no explicit spectral parameter dependence
(see for examples, \cite{Q-systems} and references therein)}, relevant to the quasiclassical limit of the string sigma-model, in terms of characters of  specific infinite-dimensional representations of \(U(2,2|4)\), parameterized by 8 eigenvalues of an arbitrary group element.  We will also demonstrate this construction in the asymptotic, large size  limit (of long SYM operators) when the asymptotic Bethe ansatz (ABA) \cite{Beisert:2005fw,Beisert:2006ez} is applicable. We will also discuss the analyticity and symmetry properties of these \(\Qs\)-functions.  The relation of the character solution to the quasi-classical limit of the superstring on the  \(AdS_5\times S^5\) background discussed in \cite{Gromov:2010vb}  will serve as an important source of inspiration for that.

 The  T-system first appeared in the context  of the quantum integrable systems
for the \(gl(2)\) algebra in \cite{Pearce:1991ty}, and generalized to \(gl(N)\) in \cite{Kuniba:1993cn}. Wronskian determinant solutions of T-system were introduced  in \cite{Bazhanov:1996dr} for \(N=2\) and
 in \cite{Krichever:1996qd} for any \(N\), where  the finite dimensional representations of \(gl(N)\) symmetry impose  the boundary conditions in the semi-infinite strip of the width \(N\) in the \((a,s)\)-representation space.
The Wronskian solutions were generalized  to the
 supersymmetric \(gl(M|N)\) algebras for  the  \(T\)-system\footnote{
T-system for this case appeared in \cite{Tsuboi:1997iq,Tsuboi:1998ne}.
}
 with
 the \((M|N)\) fat hook boundary conditions
for \((M,N)=(2,1)\) in \cite{Belitsky:2006cp,Bazhanov:2008yc} and
for any \((M,N)\) in \cite{Tsuboi:2009ud}.
On the way to  constructing similar solutions  for  the \(\mathbb{T}\)-hook boundary conditions corresponding to the superconformal \(PSU(2,2|4)\) symmetry of the model we
 will first remind the ``classical'' solution for the \(U(2,2|4) \) characters \cite{Gromov:2009tq,Gromov:2010vb} and then    discuss the most general form of the so called TQ-relations relating T- and \(\Qs\)-functions  (analogue of Baxter's famous relation for the XXZ model), in the form of the so called generating functional  \cite{Krichever:1996qd}, for finite  \cite{Tsuboi:1997iq,Tsuboi:1998ne,Kazakov:2007fy},  and even for the infinite-dimensional \cite{Beisert:2005di}
representations of  \(gl(M|N) \), as well as the
QQ-relations among the \(\Qs\)-functions, especially efficient for the super-algebras  \cite{GS03,Kazakov:2007fy}
(see also earlier papers \cite{Woynarovich83,Tsuboi:1998ne}).
The B\"acklund solution of Hirota equation for the \(gl(M)\)  algebras   \cite{Krichever:1996qd}  and \(gl(M|N)\) super-algebras, for the   fat hook boundary conditions in the representational \((a,s) \) space  \cite{Kazakov:2007fy},  was  generalized to the case of a general  \(\mathbb{T}\)-hook in \cite{Hegedus:2009ky}.
The latter can be conveniently rewritten through the generating functional  \cite{Beisert:2005di,Gromov:2010vb}  further used in this paper.

The Wronskian solution  for the AdS/CFT Y-system which  we are proposing in this paper summarizes in the most explicit and concise way all these developments. Its main advantage w.r.t. the previous solutions  \cite{Hegedus:2009ky,Gromov:2010vb}    of the Y-system (and the associated T-system) in the   \(\mathbb{T}\)-hook is its absolutely explicit form in terms of the Wronskian determinants of a finite number of Baxter's Q-functions (7 independent Q-functions) which does not include any infinite sums or integral operators. We consider this Wronskian representation as a good starting point for trying to derive a FiNLIE system describing the spectrum of the planar AdS\(_5\)/CFT\(_4\).

\section{$Y$-system and $T$-system for the spectrum of  AdS/CFT}

In this section we will remind the Y-system for the spectrum of AdS\(_5\)/CFT\(_4\), point out its symmetries and discuss its analytic properties.

\begin{figure}[ht]
\begin{center}
\includegraphics[scale=0.7]{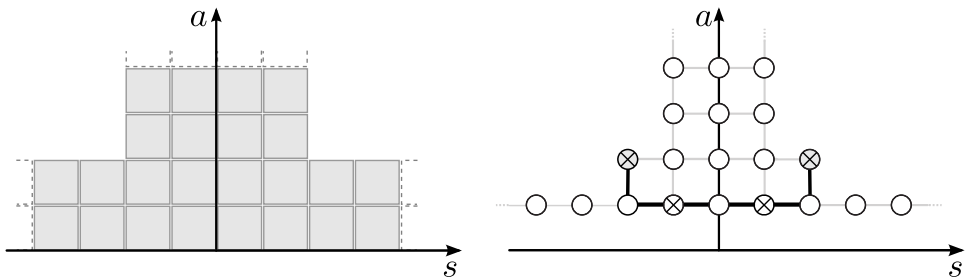}
\end{center}
\caption{\textbf{T}-shaped ``fat hook'' (\(\mathbb{T}\)-hook) uniting two \({\rm SU}(2|2)\) fat hooks,
see \cite{Gromov:2009tv} for this \(\mathbb{T}\)-hook and its generalization \cite{Hegedus:2009ky}.
}\label{T-Hook}
\end{figure}

The problem of  the spectrum of the closed superstring sigma model on the \(AdS_5\times S^5\) background, similarly to all known integrable sigma models on a space-time cylinder with the global \(gl(M|N)\) type symmetry, can be reduced to the Y-system \cite{Gromov:2009tv}

\begin{equation}
\label{eq:Ysystem} Y_{a,s}^+ Y_{a,s}^-
=\frac{(1+Y_{a,s+1})(1+Y_{a,s-1})}{(1+1/Y_{a+1,s})(1+1/Y_{a-1,s})} \,
\end{equation}
where \(Y_{a,s}\) are functions of the spectral parameter \(u\), defined   on the visible nodes of the two-dimensional integer \((a,s)\)-lattice presented on the Fig.\ref{T-Hook}(right). By the subscripts \(f^\pm\)
of any function \(f(u)\) of \(u\),
we denote the shifts of the spectral parameter by \(\pm \frac{i}{2}\): \(f^\pm\equiv f(u\pm \frac{i}{2})\).\footnote{ and in general \(f^{[\pm s]}\equiv f(u\pm \frac{is}{2})\).}

What distinguishes the Y-systems for various sigma models is: 1) the boundary conditions on the \(a,s\) lattice which are mostly defined by the symmetry of the model
2)  the analytic properties w.r.t. the spectral parameter \(u\),
only partially constrained by the symmetry and greatly depending on the physical properties of the model.

The Y-system for AdS/CFT, as any Y-system, is directly related to the so called T-system.  Namely, defining

\begin{equation}
Y_{a,s}=\frac{T_{a,s+1}T_{a,s-1}}{T_{a+1,s}T_{a-1,s}}\;
\label{eq:YTdef}\end{equation}
we rewrite \eqref{eq:Ysystem}
in terms of the bilinear finite difference    Hirota
equation
\begin{equation}
\label{fullTsystem}
T_{a,s}^+T_{a,s}^- =T_{a+1,s}T_{a-1,s}+T_{a,s+1}T_{a,s-1} \,. \\
\end{equation}

For our string sigma model the boundary conditions on the \(a,s\) lattice are constrained by the superconformal
\(PSU(2,2|4)\) symmetry \cite{Gromov:2010vb} and impose that the T-functions,
are nonzero only inside the so called
\({\mathbb T}\)-hook \cite{Gromov:2009tv}, i.e. only on the  nodes of the visible part of the  \((a,s)\)-lattice
of the Fig.\ref{T-Hook}(left). The rest of the  T-functions
are zero:\beq
T_{a,s}=0 \qquad \text{for} \qquad a<0\cup (a>2\cap |s|>2)\;.
\eeq

These boundary conditions for the functions \(T_{a,s}\) agree with the
above mentioned boundary conditions for  the functions \(Y_{a,s}\)  of the
Fig.\ref{T-Hook}(right) if we take Y-functions to be zero on the  vertical boundaries, and to be infinite on the horizontal boundaries on the
Fig.\ref{T-Hook}(left). Note that this leaves an ambiguity about the values at the corner nodes \(Y_{2,\pm 2}\).  A more careful analysis shows that
equations for \(Y_{2,\pm 2}\) are more complicated and cannot be written in a ``local'' functional form. The two missing equations \eqref{eq:Y11Y22}
do not have the standard Y-system form and can be borrowed  from the TBA approach  \cite{Gromov:2009bc}.
However, we believe that all the  functions  \(T_{a,s}\)  satisfy the standard Hirota equation \eq{fullTsystem} \cite{Gromov:2009tv}. In that sense, the T-system looks more universal, and in many cases more convenient than the Y-system.

Notice that the parameterization in terms of \(T_{a,s}\) is not unique.
The T-system is invariant under the  gauge transformation:
\begin{equation}\label{eq:GaugeTr}
T_{a,s}\to g_1^{[a+s]} g_2^{[a-s]} g_3^{[s-a]} g_4^{[-a-s]} T_{a,s}\end{equation}
 and thus the T-functions are defined up to 4 arbitrary gauge functions \(g_{a}(u)\).
The physical quantities are computed in terms of the gauge invariant  \(Y\)-functions.

The analytic structure of the AdS/CFT Y-system is inherited to a great extent from the dispersion relations for the  elementary physical excitations -- the magnons on the infinite  spin chain representing a SYM  operator, or its AdS dual on the string side. The dispersion relation between  the energy and the momentum of such solitary excitations \cite{Santambrogio:2002sb}  is conveniently parameterized  \cite{Beisert:2004hm}   by the so called  Zhukovsky map \(\frac{u}{g}=x+1/x\), where   \(g\) is related to the
`t~Hooft's coupling \(\l=g_{YM}N_c^2\) as  \(g=\frac{\sqrt\lambda}{4\pi}\):
\begin{equation}p(u)= \frac{1}{i}\log\left( \frac{x^{+}(u)}{x^{-}(u)}\right)\;\;,\;\; \eps(u)=  1+\frac{2ig}{x^{+}(u)}-\frac{2ig}{x^{-}(u)}
\;.
\end{equation}
The inverse map is double valued and we have to distinguish  two branches -- physical and mirror (related to the exchange of time and space coordinates on the world sheet cylinder \cite{Ambjorn:2005wa,Arutyunov:2007tc}):
\begin{equation}\la{xxs}
\qquad x^\ph(u)=\frac{1}{2}\lb \frac{u}{g}+\sqrt{\frac{u}{g}-2}\;\sqrt{\frac{u}{g}+2} \rb
\;,\qquad \mx(u)=\frac{1}{2}\lb \frac{u}{g}+i\sqrt{4-\frac{u^2}{g^2}}\rb \,.\end{equation}
In the physical branch, the finite cut, by definition, connects two branch points \(u=\pm2g\) whether as in the mirror branch the cut connecting them passes through the infinity.

The \(u\)-parameterization is distinguished by the fact that the fusion of elementary excitations into various bound states is especially simple in the complex \(u\)-plane: the rapidities of the constituents of a bound state are spaced by the integers of \(i\). For example, the bound states for the energy carrying magnons mentioned above have the energy and momentum \cite{Dorey:2006dq}
\beq
\exp p_a(u)={\cal F}_a\circ \exp p(u),\qquad \e_a(u)={\cal
  F}_a\circ \e(u)
\eeq
where we defined the fusion operator
\beq
\label{eq:fusdef}
{\cal F}_a\circ f(u)=\left\{\begin{array}{cl}
\prod\limits_{j=-\frac{|a|-1}2}^{\frac{|a|-1}2}f(u+ij)\;\;& ,\;\;a>0\\
1 \;\;&,\;\;a=0 \\
\prod\limits_{j=-\frac{|a|-1}2}^{\frac{|a|-1}2}1/f(u+ij)\;\;&,\;\;a<0
\end{array}\right.\;.
\eeq
This gives for the dispersion relation of the bound states

\begin{equation}\label{eq:fusedpE}p_a(u)= \frac{1}{i}\log\left( \frac{x^{[+a]}}{x^{[-a]}}\right),\qquad \eps_a(u)= a+
 \frac{2ig}{x^{[+a]}}-
\frac{2ig}{x^{[-a]}}
\;.
\end{equation}

The Y-functions as functions
of \(u\) should inherit the multi-valuedness of the map \eq{xxs}.
Most of our experience on their analyticity properties comes from the asymptotic Bethe ansatz (ABA) \cite{Beisert:2005fw,Beisert:2006ez} corresponding to the limit   of very   long operators \(L\to\infty\) and from TBA equations for the excited states  \cite{Bombardelli:2009ns,Gromov:2009bc,Arutyunov:2009ur}.
The ABA limit of Y-system found in \cite{Gromov:2009tv}
shows that the Y-functions have branch points at \(u=\pm 2g+\frac{in}{2}\) for various \(n\)'s
(\(n \in {\mathbb Z}\)).
Similarly to the above definition of \(x^\mir\) we define the ``mirror''  sheet of  the \(Y\)-functions
with the cuts going through infinity, parallel to the real axis. We
can find    from the study of the  ABA limit (see the section
\ref{sec:ABA}) that
\(\left.Y_{1, s}\right|_{s\geq 2}\) has  4 branch cuts at \({\rm Im}\;u=\tfrac{\pm
  s\pm1}{2}\), while \(\left.Y_{a,\pm 1}\right|_{a\geq 2}\) has   4
branch cuts at \({\rm Im}\;u=\tfrac{\pm a\pm1}{2}\), and \(Y_{1,\pm
  1},\;Y_{2,\pm 2}\) have 3 cuts at  \({\rm Im}\;u=0,\pm 1\). At
finite size other cuts can  appear,
but we expect the following analyticity conditions  to hold anyway \cite{GromovKazakovReview}:
\begin{enumerate}
\item \(Y_{1,\pm s}\) have no branch cuts inside the strip \(-\tfrac{|s|-1}{2}<{\rm Im}\;u<\tfrac{|s|-1}{2}\);
\item \(Y_{a,\pm 1}\) have no branch cuts inside the strip \(-\tfrac{a-1}{2}<{\rm Im}\;u<\tfrac{a-1}{2}\);
\item \(Y_{a,0}\) have no branch cuts inside the strip \(-\tfrac{a}{2}<{\rm Im}\;u<\tfrac{a}{2}\);
\item \(Y_{1,\pm 1},\;Y_{2,\pm 2}\) have a cut on the real axes such that
\beq\label{eq:Y11Y22}
 Y_{1,\pm 1}(u+i0)Y_{2,\pm 2}(u-i0)=1\,,\qquad
u \in (-\infty,-2g]\cup[2g,\infty);
\eeq
\item \(Y_{a,s}\) obtained from the generating functional should be real functions in ``mirror'' kinematics.
\end{enumerate}

The \(Y\)-system should be satisfied for the mirror branch of \(Y_{a,s}\) described above (i.e.
for the cuts chosen to go through infinity).
We also define the ``physical'' branch of \(Y_{1,0}\) as an analytic continuation
through the first cut above the real axis and then back to the real axis
(as proposed in \cite{Gromov:2009zb}),
and denote the continued function as \(Y^\ph_{1,0}\).
These properties are very similar to the ones used in \cite{Gromov:2008gj}
to convert the \(su(2)\) principal chiral filed theory Y-system into the corresponding FiNLIE.

Once the appropriate solution of the  Y-system for a given physical
state of the \(\linebreak[1] \mathrm{AdS}_5/\mathrm{CFT}_4\) system
is found  its energy   is given
 by
the following formula
\begin{equation}
E=\sum_{j}\epsilon^{\ph}_1(u_{4,j})+\sum_{a=1}^\infty\int_{-\infty}^{\infty}\frac{du}{2\pi i}\,\,\frac{\partial\epsilon^{\mir}_a}{\partial u}\log\left(1+Y_{a,0}(u)\right)\;, \label{eq:Energy}
\end{equation}
where we also have to impose the Bethe ansatz equation for the  Bethe roots (the rapidities of physical excitations) \cite{Gromov:2009bc,Gromov:2009zb}
\begin{equation}
Y^\ph_{1,0}(u_{4,j})=-1\,.
\end{equation}
the exact Bethe equations for the auxiliary roots should come from the
condition of pole cancellation.

In the subsequent sections we first recall the solution
of the \(Y\)-system in the classical large \(\lambda\) limit
given in terms of the characters and then describe our
construction for the general quantum solution.

\section{  $SU(2,2|4)$ character solution of the Y-system and classical limit}
In the classical limit when \(\lambda\) is large the shifts by \(\pm i/2\)
in the \eq{fullTsystem} become irrelevant (see \cite{Gromov:2009tq} for more details) and the functional
\(T\)-system reduces to an algebraic set of equations called in the
mathematical literature the \(Q\)-system\footnote{It should not be confused with the Baxter's Q-functions considered below.}:\begin{equation}
\label{simpleTsystem}
\left(T_{a,s}\right)^2 =T_{a+1,s}T_{a-1,s}+T_{a,s+1}T_{a,s-1} \,.
\end{equation}

In the paper \cite{Gromov:2010vb} three of the current authors clarified  the group theoretical meaning of the AdS/CFT
\(Y\)-system, its relation to the characters of irreducible representations of the  \(SU(2,2|4)\) symmetry, related to the superconformal
\(PSU(2,2|4)\) symmetry of the model, and their relation to the classical limit,
extending some of the results of \cite{Gromov:2009tq} to all sectors. We will briefly remind in this subsection the basic results concerning the explicit construction of the
\(SU(2,2|4)\) characters for the unitary  representations, in terms of  finite determinants, in full similarity to the 1-st Weyl formula known for the compact representations of
\(GL(N)\).  In the next chapter we will show how to generalize these formulas to the quantum solution of the T-system \eq{fullTsystem},
and to find the general explicit solution of the underlying
\(Y\)-system and \(T\)-system in terms of finite determinants, called Wronskians,
parameterized by a finite number of Baxter's \(\Qs\)-functions.
We hope to apply these results in the future for the construction of a finite system
of non-linear integral equations (FiNLIE) describing the full spectrum of the planar AdS\(_5/\)CFT\(_4\).

\subsection{The character solution of the simplified AdS/CFT Y-system in $\mathbb{T}$-hook}

The generating function of   \(U(2,2|4)\) characters of ``symmetric'' representations can  be represented as
\begin{eqnarray}\label{eq:genfin}
w_{4|4}(t;h)
=\frac{(1-y_{\jA}t)(1-y_{\jB}t)}
{(1-x_{\ja}t)(1-x_{\jb}t)}\times\frac{(1-y_{\jC}t)(1-y_{\jD}t)}
{(1-x_{\jc}t)(1-x_{\jd}t)}
\label{GEN44}
\end{eqnarray}
 where \((x_\ja,\dots,x_\jd|y_\jA,\dots,y_\jD)\) are the eigenvalues of a group element \(h\in U(2,2|4)\).
 Here and below we use the hats over indexes to indicate their ``fermionic'' grading, as opposed to the ``bosonic'' grading of the rest of indices.

The first and the second factors in the r.h.s. may be attributed to the right and left \(U(2|2)_{R,L}\) subgroups.
The characters of ``symmetric'' representations are generated by  \begin{equation}
T_{1,s}^{(4|4)}[h]=\oint_{C}\frac{dt\,\, w_{4|4}(t;h) }{2\pi i}t^{-s-1}\;,
\label{oint44}\end{equation}
where the integration contour \(C\) encircle  \(t=0\) together with
the  poles \(\frac{1}{x_\jc},\frac{1}{x_\jd}\)
corresponding to the  subgroup \(U(2|2)_L\),     leaving outside
the poles  \(\frac{1}{x_\ja},\frac{1}{x_\jb}\,\) corresponding to the
first subgroup  \(U(2|2)_R\).
Note that \(s\) can here be positive as well as negative: \(-\infty<s<\infty\) and the corresponding irreps, first constructed in \cite{CLZ03,Kwon06}, are infinite-dimensional: \(T_{1,s}^{(4|4)}[h]\) are not  polynomials of \(x_i,y_\jJ\) anymore, unlike the compact representations of
\(U(M|N) \).  The rest of the \(T_{a,s}(h)\) can be restored by means of the Jacobi-Trudi type formula:
\begin{equation}\label{eq:Jacobi-Trudi}
T_{a,s}=\det_{1\le i,j\le a}\, T_{1,s+i-j}\;.
\end{equation}
It is easy to see (at least on \textit{Mathematica}, using the code given in \cite{Gromov:2010vb}) that \(T_{a,s}\ne 0\) only for \((a,s)\in\mathbb{T}\) where by  \({\mathbb T}\) we denote the \(\mathbb{T}\)-hook drawn on
Fig.\ref{T-Hook}(left). In \cite{Gromov:2010vb} the explicit expressions for  all these characters in terms of \(2\times 2\) and \(4\times 4\) determinants was found:
\begin{eqnarray}\label{CharSol}
T_{a,s}=
\left\{\bea{cc}
(-1)^{(a+1)s}\left(\frac{x_\jc x_\jd}{y_\jA y_\jB y_\jC
    y_\jD}\right)^{s-a}\frac{{\rm det}\left(S_\jI^{\theta_{j,s+2}}
    y_\jI^{j-4-(a+2)\theta_{j,s+2}}\right)_{1\le i,j\le 4}}
{
{\rm det}\left(S_\jI^{\theta_{j,0+2}}
  y_\jI^{j-4-(0+2)\theta_{j,0+2}}\right)_{1\le i,j\le 4}
}&,\;\;a\geq|s|\\
\frac{{\rm det}\left(Z_i^{(1-\theta_{j,a})} x_i^{2-j+(s-2)(1-\theta_{j,a})}\right)_{1\le i,j\le 2}}
{
{\rm det}\left(Z_i^{(1-\theta_{j,0})} x_i^{2-j+(0-2)(1-\theta_{j,0})}\right)_{1\le i,j\le 2}
}
&,\;\;s\geq +a
\eea\right.
\end{eqnarray}
where
\(
S_\jI=\frac{(y_\jI-x_\jc)(y_\jI-x_\jd)}{(y_\jI-x_\ja)(y_\jI-x_\jb)}\),
\(Z_i = \frac{(x_i-y_\jA)(x_i-y_\jB)(x_i-y_\jC)(x_i-y_\jD)}
{(x_i-x_\jc)(x_i-x_\jd)}\)
and
\(
\theta_{j,s}=\left\{\bea{cc}
1&,\;\;j>s\\
0&,\;\;j\le s
\eea\right.\;.
\)
The other \(T\)'s can be obtained using the wing-exchange symmetry
which is related to an outer automorphism of the Dynkin diagram of \(\gl(4|4)\)
\beq
 T_{a,s}(x_{\ja},\dots,x_{\jd}|y_{\jA},\dots,y_{\jD}) =
\lb\frac{y_\jA y_\jB y_\jC y_\jD}{x_\ja x_\jb x_\jc x_\jd}\rb^a T_{a,-s}\left(\left.\frac{1}{x_{\jd}},\dots,\frac{1}{x_{\ja}}\right|\frac{1}{y_{\jD}},
\dots,\frac{1}{y_{\jA}}\right).
\eeq
Another important property, under the rescaling of eigenvalues,
reads
\beq
T_{a,s}(\alpha x_i,\alpha y_\jJ)=\alpha^{a s} T_{a,s}(x_i,y_\jJ)
\label{scT}
\eeq
which implies that the gauge invariant quantities are invariant under
the P-transformation from PSU\((2,2|4)\).
 Note that the Weyl group symmetry w.r.t. permutations of
 \((x_\ja,x_\jb,x_\jc,x_\jd)\) is broken: these characters are
 symmetric only w.r.t. \(x_\ja\leftrightarrow x_\jb\) and
 \(x_\jc\leftrightarrow x_\jd\).

As we mentioned in the beginning of the section this character solution in  the \(\mathbb{T}\)-hook is tightly related to the classical limit of the  AdS\(_5\times\)CFT\(_4\) Y-system.
In this limit the string is long, the AdS/CFT coupling
is large and the natural scale of the spectral parameter is \(u\sim g\).
We see that we can neglect the shifts w.r.t. the spectral parameter in \eqref{eq:Ysystem} and \eqref{fullTsystem} (only a slow parametric dependence of the group element \(h(u)\) is left) and Hirota equations takes a simplified form, called Q-system \eqref{simpleTsystem}.
Note that \eqref{CharSol} represents the most general
solution of  this equation with the \(\mathbb{T}\)-hook boundary conditions
and is parameterized by 8 eigenvalues\footnote{One eigenvalue can be always rescaled to unity leaving us with only \(7\) gauge independent parameters.}.

It was shown in \cite{Gromov:2010vb} that the characters  \eqref{CharSol} give the classical limit of the  AdS/CFT
T-system  (or Y-system)  if the group element \(h\) is simply the monodromy matrix
\(\Omega(u)\) of  the    classical finite gap solution of Metsaev-Tseytlin superstring \cite{Bena:2003wd,Kazakov:2004qf,Beisert:2004ag,Kazakov:2004nh,Beisert:2005bm}

\begin{equation}T_{a,s}={\rm Str}_{a,s}\Omega\;.\end{equation}

Thus the solution of Hirota equation can be expressed solely in terms of the
eigenvalues of the classical monodromy matrix. The eigenvalues as functions of the spectral parameter represent 8 sheets
of the finite gap algebraic curve and the generating function \eqref{eq:genfin} can be viewed as its spectral super-determinant.

\subsection{Quantization of the classical algebraic curve}\label{sec:AlgCurve}
\begin{figure}[ht]
\begin{center}
\includegraphics[scale=1]{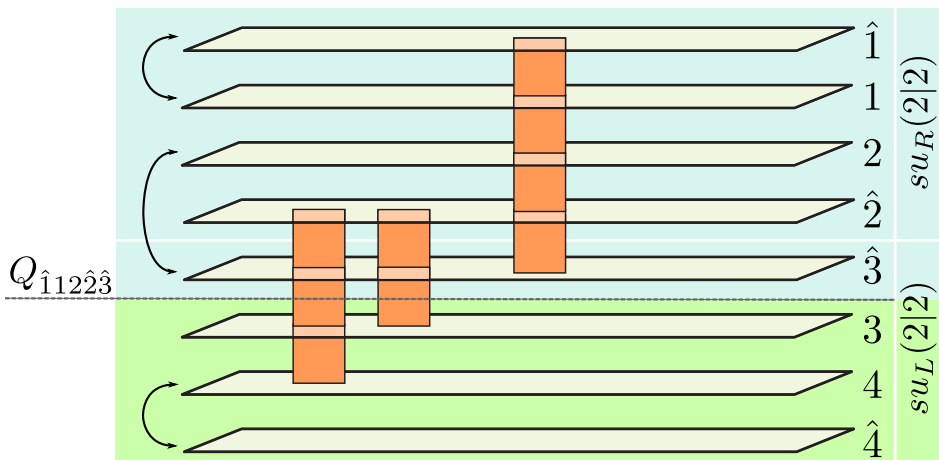}
\end{center}
\caption{One can quantize the classical curve by choosing  different orders of the sheets. Once an
order (called path) is fixed the \(\Qs\) functions will correspond to the cuts crossing the line
between two neighbor sheets. A \(\Qs\) function is not affected by the permutations
 of sheets above and below the line. Thus the \(\Qs\) function
 is uniquely determined by the subset of the sheets above the line and we denote it
as \(Q_{\jA\ja\jb\jB\jC}\).
For more formal definitions see section 4.3.
}\label{fig:dyal}
\end{figure}

For the states with large length \(L\) the spectral equations should simplify to ABA equations. These states
are described within the ABA approach  by the configurations of the Bethe roots.
A map between the configurations of the Bethe roots and the classical algebraic curve was understood  in  \cite{Kazakov:2004qf,Arutyunov:2004vx,Beisert:2005di,Beisert:2005fw,Gromov:2006dh,Gromov:2006cq}.
The basic idea is that the branch cuts of the classical curve are the cuts on the
complex plane where the Bethe roots are densely distributed. Technically one can write
the classical quasi-momenta in terms of the densities of the Bethe roots
or equivalently resolvents:
\beq
H_a=\sum_{j=1}^{K_a}\frac{x^2}{x^2-1}\frac{1}{x-x_{a,j}}\;\;,\;\;\bar
H_a(x)=H_a(1/x)\;.
\eeq
where \(a=1,\dots,7\) is a type of the Bethe root.
The   eigenvalues of the monodromy matrix  are related to the quasi-momenta by the formulas (see the notations in   \cite{Beisert:2005bm,Gromov:2007cd})

\beqa
\bea{l}
y_\jA =e^{-i\hat p_1}= \exp\lb- i\frac{L x/(2g)  -i\mathcal{Q}_2 x}{x^2-1} -i H_1 -i \bar H_3 +i \bar H_4\rb\\
x_\ja =e^{-i p_1}= \exp\lb-i {\frac{L x/(2g) +i \mathcal{Q}_1}{x^2-1}} -i H_1 +i H_2 +i \bar H_2 -i \bar H_3\rb \\
x_\jb =e^{-i p_2}= \exp\lb-i {\frac{L x/(2g)  +i \mathcal{Q}_1}{x^2-1}} -i H_2 +i H_3 +i \bar H_1 -i \bar H_2 \rb\\
y_\jB =e^{-i\hat p_2}= \exp\lb- i {\frac{L x/(2g)  -i \mathcal{Q}_2x }{x^2-1}} +i H_3 -i H_4 +i \bar H_1 \rb \\
y_\jC =e^{-i\hat p_3}= \exp\lb+ i{\frac{L x/(2g) -i \mathcal{Q}_2x }{x^2-1}}  -i H_5 +i H_4 -i \bar H_7 \rb\\
x_\jc =e^{-i p_3}= \exp\lb+i {\frac{L x/(2g) +i \mathcal{Q}_1}{x^2-1}} +i H_6-i H_5 -i \bar H_7+i \bar H_6 \rb \\
x_\jd =e^{-i p_4}= \exp\lb+i {\frac{L x/(2g)  +i\mathcal{Q}_1}{x^2-1}}  +i H_7-i H_6 -i\bar H_6 +i \bar H_5\rb \\
y_\jD =e^{-i\hat p_4}= \exp\lb+i {\frac{L x/(2g) +i \mathcal{Q}_2 x }{x^2-1}} +i H_7 +i \bar H_5 -i \bar H_4\rb\,\,.
\eea
\label{eq:p}
\eeqa
Where \({\cal Q}_1=\sum_j\frac{4\pi x_4}{\sqrt\lambda(x_4^2-1)}\),
\({\cal Q}_2=\sum_j\frac{4\pi}{\sqrt\lambda(x_4^2-1)}\).
Notice that only two subsequent sheets share the same
resolvent.
This discretization is not unique. In particular, a configuration of Bethe roots could be mapped to a
different one using the so-called fermionic and bosonic dualities, which correspond to a
reshuffling of the sheets of the Riemann surface (e.g., see \cite{Gromov:2010vb}).   One can make a permutation
of  sheets and discretize the quasi-momenta in terms of a new set of Bethe roots defining  new resolvents
(see Fig.\ref{fig:dyal}).  The cuts connecting some non-neighboring sheets  can cross on the way the other sheets due to a mechanism of
``stuck formation'' of various types of Bethe roots \cite{Beisert:2005di} (strictly speaking the stacks are only formed for
small enough filling fractions and nonzero twists \cite{Gromov:2007ky}).

We will see that it is useful not to restrict ourselves to a single possible discretization (related to the quasi-classical  quantization) of the curve
  but rather to consider all possible permutations of the sheets.
These permutations can be denoted by  sequences of indices labeling the sheets.  In these notations, we denote the chosen ordering (grading) of sheets as \(\jA\ja\jb\jB\jC\jc\jd\jD\).

The grading   \(\jA\ja\jb\jB\jC\jc\jd\jD\) has some obvious advantages in the ABA limit and we will
use it as a canonical one.

The main motivation of this paper is the generalization of this solution to the full quantum case of the AdS\(_5/\)CFT\(_4\) Y-system valid for any operator and any \(\l\). In the next section we show that at least the first part of this program
-- the general solution of the T-system \eqref{fullTsystem} in the \(\mathbb{T}\)-hook, in terms of  a finite number of functions of \(u\)
--  the Baxter's \(\Qs\)-functions -- can be fulfilled in a rather explicit form.
The next step -- finding the general analyticity properties of these functions and constructing the corresponding
FiNLIE, will be the subject of a future work.

\section{Solution of the full Hirota equation in  $\mathbb{T}$-hook}

Our goal here will be the construction of a general solution of Hirota equation \eqref{fullTsystem} for the \((2|4|2) \)    \(\mathbb{T}\)-hook, first in terms of the generating functional (GF) and then in explicit and finite form, in terms of the
Wronskians --  the finite determinants of \(\Qs\)-functions.

\subsection{Generating functional for the $\mathbb{T}$-hook}\label{sec:GenFun}

For given \(T_{1,s}(u),\quad-\infty<s<\infty\), the general solution of \eqref{fullTsystem}  is given by the Bazhanov-Reshetikhin  (BR) type formula\footnote{The original  BR formula was written for the fusion in spin chains.  It is a generalization of the Jacobi-Trudi  formula \eqref{eq:Jacobi-Trudi} for the characters. Here we view \eqref{eq:BR} as a general solution of Hirota equation  on the upper part of the  \((a,s)\)-lattice, with the Dirichlet boundary conditions: \(T_{0,s}=1\) and \(T_{1,s}\) -- fixed. }
 \begin{equation}\label{eq:BR}
T_{a,s}=\det_{1\le j,k\le a}T_{1,s+j-k}^{[-a+k+j-1]}\;.
\end{equation}

As was shown in  \cite{Tsuboi:1997iq,Tsuboi:1998ne,Kazakov:2007fy}, to solve Hirota equation in  a \((M|N)\) fat hook the ``symmetric'' T-functions \(T_{1,s}(u) \) should be generated by a special generating functional. In particular, for the  \((2|2)\) fat hook, in the grading \((1|2|1)\) corresponding to the Kac-Dynkin diagram
\(\otimes\!\!-\!\!\odot\!\!-\!\!\otimes\), or to the \(\{y_{\jA}|x_1,x_2|y_{\jB}\}\)      ordering of the eigenvalues of a twist parameter \(g\in GL(2|2)\), we have the following generating functional
\begin{eqnarray}\label{eq:GenFn22}
W=\Uh_{,\jA}\,\Uh_{\jA,\ja}^{-1}\,\Uh_{\jA\ja,\jb}^{-1}\,\Uh_{\jA\ja\jb,\jB}=\sum_{s=-\infty}^{\infty} D^s T_{1,s}^{}
D^{s}
\end{eqnarray}
where
\begin{equation}\label{eq:Mon-chi}
\Uh_{A}=(1-D\chi_{A}D),\qquad \qquad (D=e^{-\frac{i}{2}\p_u})
\end{equation}
with \(A\subset I_0= \{\jA|\ja,\jb|\jB\}\)  being an arbitrary  subset
of the full set \(I_0\), while\\
  \(\{\chi_{,\jA}(u)|\chi_{\jA,\ja}(u) ,
  \chi_{\jA\ja,\jb}(u)|\chi_{\jA\ja\jb,\jB}(u)\}\)
  are  arbitrary functions of the spectral parameter \(u\),
  replacing the group element eigenvalues
  \(\{y_{\jA}|x_{\ja},x_{\jb}|y_{\jB}\}\) of the character generating
  function \eqref{GEN44}, whereas the generating parameter \(t\) is
  replaced by the shift operator \(D=e^{-\i2\p_u}\).  As was proposed
  in \cite{Tsuboi:2009ud} in the case of a more general
superalgebra\footnote{It is important to note that the Baxter \(\Qs\)-operators
are essentially independent of the shape of the ``hook''.
They are fixed by a certain oscillator  algebra \cite{Bazhanov:1996dr,Bazhanov:1998dq,Bazhanov:2001xm,Bazhanov:2008yc,BFLMStoappear},
at least
for the models with the Yangian or the quantum affine algebra symmetries.
This allows to apply
 the formalism for the \((M|N)\)-hook
 directly to the construction of the Wronskian solution for
the \(\mathbb{T}\)-hook in question.
}
\(gl(M|N)\) (and demonstrated on the Fig.\ref{fig:dyal}), the labeling here
  follows the  pattern of the Dynkin diagram in the grading
  \(\otimes\!\!-\!\!\odot\!\!-\!\!\otimes\):  we start from the
  l.h.s. of this diagram with \(A=(,\jA)\) for the
  first factor in the
  l.h.s. of \eqref{eq:GenFn22}; then for the label \(A\) of the second
  factor, one moves the comma one step to the right and adds
  new index after the comma, making \(A=(\jA,i)\). The choice of the
  index is related to the grading: when  moving
  to the right of the Dynkin diagram we cross
  the fermionic node  which means that we  chose to add an entry from \(I\)
  with a different grading than the previous \(i\), say \(\ja\),
  which gives for the second factor \(A=(\jA,\ja\)); then we cross the
  bosonic node, which means that we should add an entree with the same
  grading as the last one, i.e. \(\jb\), which gives for the label of the third factor
  \(A=(\jA\ja,\jb)\), and finally, crossing the last fermionic node
  we get for the label of the last factor \(A=(\jA\ja\jb,\jB)\).  This
  means that  the subset
  \(A\) represents a ``path'' by which it was reached starting from
  the l.h.s. of the Dynkin diagram. This will be the general rule of
  for more complicated algebras, and in particular
  \(u(2,2|4)\).\footnote{\label{foot:Order}
We will see later that not all the paths give inequivalent \(\chi_{A,i}\)
functions. Namely, \(\chi_{I,i}=\chi_{I',i'}\) iff  \(I\)
and \(I'\) differ only by the permutation of their indices, while \(i=i'\).}

To calculate \(T_{a,s}(u)\) in terms of these functions we formally expand the l.h.s. in powers of the shift operator \(D=e^{-\i2\p_u}\) and compare the coefficients with the r.h.s.
The formal proof can be found in
\cite{Tsuboi:1997iq,Tsuboi:1998ne,Kazakov:2007fy},
but it is easy to convince oneself on \textit{Mathematica} that the \(T\)-functions generated from \(T_{1,s}\) by  \eqref{eq:BR} are zero outside of the \((2|2)\) fat hook.

What is the solution for the  \(\mathbb{T}\)-hook of \(u(2,2|4)\)? We can read it off from the generating function  for the characters of  \(u(2,2|4)\)  given by eqs.\eqref{GEN44}-\eqref{oint44}. Note that the integration contour prescription in \eqref{oint44} means that we can generate the  \(u(2,2|4)\) characters by expanding the first factor in the r.h.s. of \eqref{GEN44} in powers of \(t\), and the second factor -- in powers of \(\frac{1}{t}\), and then extract from the product   \(T_{1,s}\) as a coefficient in front of the power \(t^s\). The result is given in terms of infinite sums, and not polynomials in the eigenvalues, signalling that we deal with infinite-dimensional irreps of \(u(2,2|4)\) \cite{CLZ03,Beisert:2005di,Kwon06}.

Similarly   to the case \(gl(2|2) \) described above, we can try as a  quantum generalization of equation \eqref{fullTsystem} for the  \((2|4|2)\) \(\mathbb{T}\)-hook,
for the particular grading   of the eigenvalues \(\{y_\jA|x_\ja,x_\jb|y_\jB,y_\jC|x_\jc,x_\jd|y_\jD\}\) corresponding to the Kac-Dynkin diagram \(\otimes\!\!-\!\!\odot\!\!-\!\!\otimes\!\!-\!\!\odot\!\!-\!\!\otimes\!\!-\!\!\odot\!\!-\!\!\otimes\), the following generating functional
\cite{Beisert:2005di}
\begin{eqnarray}\label{eq:GenFn224}
W&=&\left[\tU_\jA\,\tU_{\ja}^{-1}\,\tU_{\jb}^{-1}\,\tU_{\jB}\right]_{+}\times
\left[\tU_{\jC}\,\tU_{\jc}^{-1}\,\tU_{\jd}^{-1}\,\tU_{\jD}\right]_{-}
\!\!=\!\!\sum_{s=-\infty}^{\infty} D^{s}T_{1,s}^{}D^{s}
\end{eqnarray}
where we introduced, to make the formula less bulky,  the  notations
\(\tU_j=1-D\xi_j D\) with
\begin{eqnarray}\label{eq:GF224notations}
\xi_{\jA}&=&\chi_\jA\,
\qquad\quad\,\,\,\xi_{\ja}=\chi_{\jA,\ja}\,
\qquad\quad\xi_\jb=\chi_{\jA\ja,\jb}\,
\qquad\xi_\jB=\chi_{\jA\ja\jb,\jB}\,
\nn\\
\xi_{\jC}&=&\chi_{\jA\ja\jb\jB,\jC}\,
\qquad\xi_\jc=\chi_{\jA\ja\jb\jB\jC,\jc}\,
\quad\xi_{\jd}=\chi_{\jA\ja\jb\jB\jC\jc,\jd}\,
\quad\xi_\jD=\chi_{\jA\ja\jb\jB\jC\jc\jd,\jD}\;.
\end{eqnarray}
We label the functions \(\xi_j\) only by the last index in the subset \(A\) of  \(\chi_{A}\) having in mind  the particular grading, or nesting path. By definition, we expand   inside  \(\left[\dots\right]_+\)
and   \(\left[\dots\right]_-\) w.r.t. the positive and negative  powers of \(D\) respectively and then calculate the functions \(T_{1,s},\,\,-\infty<s<\infty\), comparing the powers of \(D\) on both sides of the eq.\eq{eq:GenFn224}. To restore the rest of \(T_{a,s}\) we can use again \eqref{eq:BR}.

The formal proof\footnote{ The B\"acklund procedure of  \cite{Kazakov:2007fy,Hegedus:2009ky} can be used for it.
}
of the fact that \eqref{eq:BR},\eqref{eq:GenFn224}  represents the complete solution of Hirota equation within the \(\mathbb{T}\)-hook will be published elsewhere, but, again,  it is easy to convince oneself in the correctness of this formula on \textit{Mathematica}.

The representation \eqref{eq:GenFn224} is already  a considerable advance w.r.t. the original Y-system \eqref{eq:Ysystem} since its solution is now parameterized in terms of a finite number of functions
\(\xi_j\).
A solution of Hirota equation is now parameterized by   \(8\) arbitrary independent functions \(\xi_j\) entering the generating functional \eq{eq:GenFn224}\footnote{One of them can be removed by a gauge.}.
Moreover in this form the passage to the general strong coupling solution found in \cite{Gromov:2010vb} is straightforward -- in the strong coupling limit the shift
operator \(D\) can be replaced by a formal scalar expansion parameter \(t\) whereas \(\xi_j\) should become the eigenvalues of the classical monodromy matrix.
This solution is quite useful for  various applications. Note however that as the result we get infinite series, signaling that we deal with  infinite dimensional  representations of \(U(2,2|4)\). In the next section, we will show that these infinite series can be
``miraculously'' converted to the explicit finite dimensional Wronskian determinants.

\subsection{QQ-relations}

To find the Wronskian solution we have to choose a good parameterization for all these 8 functions \eq{eq:GF224notations}. The formalism of
\cite{Tsuboi:1997iq,Tsuboi:1998ne,Beisert:2005di,Kazakov:2007fy,Tsuboi:2009ud}
aimed at the derivation of  Bethe equations as a condition of analyticity, i.e. polynomiality, of the T-functions, as well as the form of the classical finite gap solution \eqref{eq:p} suggests that the best parameterization would be in terms of the \(\Qs\)-functions (a-la Baxter).  As it was explained in sections \ref{sec:AlgCurve} and \ref{sec:GenFun} and on the Fig.\ref{fig:dyal}, all the \(2^{8}\) \(\Qs\)-functions
can be labeled by  all possible  subsets
from the full set \(I_0=\{\ja,\jb,\jc,\jd|\jA,\jB,\jC,\jD\}\).

\begin{figure}
  \centering
  \includegraphics{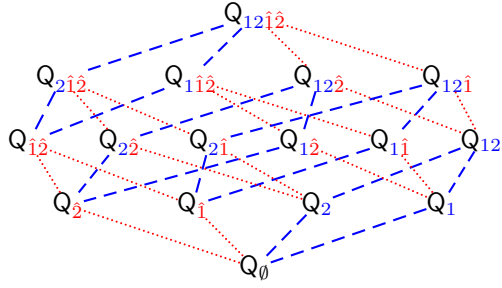}
  \caption{
\textbf{Hasse diagram for \(gl(2|2)\) (cf.\ \cite{Tsuboi:2009ud}):}
To construct the generating functional \protect\eqref{eq:GenFn22} we move along any path starting at the node \(\emptyset\) and ending at the node \(\ja\jb\jA\jB\). Each line corresponds  to adding the next \(U\)-factor (or \(\tU\)-factor in \protect\eqref{eq:GenFn224} for the \((2,2|4)\) case) with  one extra index. The dashed blue (resp. dotted red) lines stand for adding  a ``bosonic'' (resp. fermionic) index, and each local change of the nesting path at some rectangular facet of the diagram  gives rise to the \(QQ\)-relations
\protect\eqref{eq:QQb}, \protect\eqref{eq:QQf}.
The equation \protect\eqref{eq:QQf} (resp. \protect\eqref{eq:QQb}) corresponds to the facets
having \(2\) ``bosonic'' lines and \(2\) ``fermionic'' lines
(resp to the facets having all \(4\) ``bosonic'' or all \(4\) ``fermionic'' lines).}
 \label{Hasse}
\end{figure}

The  monomials of the generating functional can  be conveniently parameterized in terms of 8 Baxter's \(\Qs\)-functions as follows:
\beq
\label{eq:chi-Q}
\bea{llllll}
\xi_{\jA} =
\frac{\Qs_{\emptyset}^{++} \Qs_{\jA}^{-}}{\Qs_{\emptyset}  \Qs_{\jA}^{+}}
\;,
&
\xi_{\ja} =
\frac{\Qs_{\jA}^{-} \Qs_{\jA\ja}^{++} }{\Qs_{\jA}^{+} \Qs_{\jA\ja} }
\;,
&
\xi_{\jb} =
\frac{\Qs_{\jA\ja}^{--} \Qs_{\jA\ja\jb}^{+} }{\Qs_{\jA\ja}  \Qs_{\jA\ja\jb}^{-}}
\;,
&\xi_{\jB} =
\frac{\Qs_{\jA\ja\jb}^{+} \Qs_{\jA\ja\jb\jB}^{--} }{\Qs_{\jA\ja\jb}^{-} \Qs_{\jA\ja\jb\jB} }
\;,\\
{\xi}_{\jD} =
\frac{\Qs_{\jA\ja\jb\jB\jC\jc\jd\jD}^{--} \Qs_{\jA\ja\jb\jB\jC\jc\jd}^{+}}{ \Qs_{\jA\ja\jb\jB\jC\jc\jd\jD} \Qs_{\jA\ja\jb\jB\jC\jc\jd}^{-}}\;,
&
{\xi}_{\jd}  =
\frac{\Qs_{\jA\ja\jb\jB\jC\jc\jd}^{+} \Qs_{\jA\ja\jb\jB\jC\jc}^{--} }{\Qs_{\jA\ja\jb\jB\jC\jc\jd}^{-}\Qs_{\jA\ja\jb\jB\jC\jc} }
\;,
&
{\xi}_{\jc} =
\frac{ \Qs_{\jA\ja\jb\jB\jC\jc}^{++} \Qs_{\jA\ja\jb\jB\jC}^{-}}{ \Qs_{\jA\ja\jb\jB\jC\jc} \Qs_{\jA\ja\jb\jB\jC}^{+}}
\;,
&{\xi}_{\jC} =
\frac{\Qs_{\jA\ja\jb\jB\jC}^{-}\Qs_{\jA\ja\jb\jB}^{++}  }{  \Qs_{\jA\ja\jb\jB\jC}^{+}\Qs_{\jA\ja\jb\jB}}
\;.
\eea
\eeq
The way we perform this parameterization for the functions \(\chi_{A}\)  (see the definition \eq{eq:GF224notations})
in terms of the
\(\Qs\)-functions is completely defined by the chosen nesting (grading)  path
and   is simply given by
 \(\chi_{I,j}=
\frac{\Qs_{I}^{[k+2\s]}}{\Qs_{I}^{[k]}}\frac{  \Qs_{I,j}^{[k-\s]}}{
  \Qs_{I,j}^{[k+\s]}}\)   where  \(\Qs_I^{[k]}\) in the denominator is
the same as the last\footnote{To understand what is the first and the
  last factor, the second line in \eqref{eq:chi-Q} should be read from
  right to left,
  which makes the number of indices increase continuously.
  } \(\Qs\) in the denominator of  \(\chi_I\)\footnote{which helps to
    cancel the  poles given by zeros of Q-functions, if there are any, in all T-functions by
    imposing the Bethe ansatz equations on their positions.
},
\(\s=\pm1\) is chosen so that the first ratio  in  \(\chi_{I,j} \) is
the same as the last ratio in \(\chi_I\) when   we change the grading
passing from \((I,i)\) to  \((Ii,j)\), and \(\s\) is opposite otherwise.
All  such nesting paths for the case of \(su(2|2)\) are shown on
the Hasse diagram in  Fig.\ref{Hasse}.

All that means that the labeling of \(\Qs\)-functions follows the nesting path.
However, it is clear from the Fig.\ref{fig:dyal} and the explanations in sec.\ref{sec:AlgCurve}  that these \(Q\)-functions do not
depend on the order of indices in the path  up to a sign: a
permutation of indices
multiplies the \(\Qs\)-function by the signature of the
permutation\footnote{\label{foot:OrderQ} The \(\pm\) sign coming
  form this signature is
irrelevant in the \(\chi_{A,i}\) functions, where these signs are
canceled in the ratios of \(\Qs\) functions. The statement of the footnote
  \ref{foot:Order} is a consequence of that.}.

Moreover, all these \(2^8=256 \) \(\Qs\)-functions are not independent and can be expressed through a  chosen basis of 8 \(\Qs\)-functions. The rest of them are related to those 8 by the so called QQ-relations
\cite{GS03,Kazakov:2007fy,QQ-boson,Bazhanov:2001xm,Belitsky:2006cp,Gromov:2007ky,Bazhanov:2008yc}:
\footnote{
Of course, any linear combination of  solutions of the Baxter equation
is also a solution. Besides that, we have a rotational symmetry \(GL(4)\times GL(4)\)
of QQ-relations.
In addition, there are discrete symmetries related to the outer automorphisms of \(gl(4|4)\)
and the gauge symmetry.
However, once we
fix a basic set of 8 arbitrary \(\Qs\)-functions, we will fix the
\(2^{4+4}\) \(\Qs\)-functions unambiguously, through the QQ-relations.
}
\begin{equation} \label{eq:QQb}
\Qs_{I}\Qs_{I,ij}
=\Qs_{I,i}^{[\ps_{i}]}
\Qs_{I,j}^{[-\ps_{i}]}-
\Qs_{I,i}^{[-\ps_{i}]}
\Qs_{I,j}^{[\ps_{i}]}
\qquad
\text{for} \qquad \ps_{i}=\ps_{j},  \\[6pt]
\end{equation}
\begin{equation} \label{eq:QQf}
\Qs_{I,i}\Qs_{I,j}=
\Qs_{I}^{[-\ps_{i}]}
\Qs_{I,ij}^{[\ps_{i}]}-
\Qs_{I}^{[\ps_{i}]}
\Qs_{I,ij}^{[-\ps_{i}]}
\qquad \text{for} \qquad \ps_{i}=-\ps_{j}.
\end{equation}
\begin{equation}
  \textrm{where}\qquad \ps_i=+1,\qquad i\in\{\ja,\jb,\jc,\jd\}\qquad\textrm{and}\qquad \ps_{\jI}=-1,\qquad \jI\in\{\jA,\jB,\jC,\jD\}
\end{equation}

These QQ-relations can be obtained from the fact that the relabeling of \(\chi\)-functions and \(\Qs\)-functions  in \eqref{eq:GenFn224} by changing the nesting path does not change the generating functional.  This change occurs from the commutation of two consecutive \(U\)-factors in the generating functional. Following the nesting path \( (I,i)\to(I,i,j) \) should give the same generating functional in \vref{eq:GenFn224} as  the path  \( (I,j)\to(I,j,i) \):
\begin{equation}
 \tU_{I,i}^{-\ps_{i}}\tU_{I,ij}^{-\ps_{j}}=\tU_{I,j}^{-\ps_{j}}\tU_{I,ji}^{-\ps_{i}}
\end{equation}
which gives both ``bosonic'' and ``fermionic'' QQ-relations \eqref{eq:QQb} and \eqref{eq:QQf}.

Let us note that the dependence on the general (\(u\)-independent) twist matrix\\ \(g={\rm diag} \{z_\jA|z_\ja,z_\jb|z_\jB,z_\jC|z_\jc,z_\jd|z_\jD\}\in GL(4|4)  \), useful for some applications, such as the \(\b\)-deformed version of the AdS\(_5\)/CFT\(_4\) duality, can be easily introduced into the \(\mathbb{T}\)-hook solution \eqref{eq:GenFn224} by a certain simple rescaling of all \(\Qs\)-functions.\footnote{
One can recover the twist \(\{z_{i}\}\) after the following  formal replacement:
\(\Qs^{[0]}_{I} \to a_{I} f_{I}^{[0]}\Qs^{[0]}_{I}\),
\(\Ts_{a,s}^{[0]} \to a_{1234 \jA \jB \jC \jD} f^{[a-s]}_{1234\jA \jB \jC \jD} \Ts_{a,s}^{[0]}\),
where \(f_{I}^{[0]}=\prod_{i \in I}f_{i}^{[0]}\),
\(f_{i}^{[s]}=z_{i}^{\frac{p_{i}s}{2}}f_{i}^{[0]}\),
\(a_{I}=\prod_{j,k \in I; j<k }
(\frac{z_{j}-z_{k}}{(z_{j}z_{k})^{\frac{1}{2}}})^{p_{j}p_{k}} \)
(cf. eqs.\ (3.26)-(3.34) in \cite{Tsuboi:2009ud}).
To take a limit \(z_{i} \to 1\) is a non-trivial task, where one have to take into account
the rotational symmetry of the QQ-relations (see recent papers
\cite{Bazhanov:2010ts}).
This produces an extra factor of \(z_j\) in the r.h.s. of each of the eight expressions  for \(\xi_j\) in \eqref{eq:chi-Q}.
}

\subsection{Explicit Wronskian formulae for $\mathbb{T}$-hook  }

Now we can formulate the central result of this paper -- the
determinant formulae for the complete solution  of the  Hirota
equation (\(T\)-system) \eqref{fullTsystem} within the \((2,2|4)\)
\(\mathbb{T}\)-hook.
They are written in a gauge where \(\Qs_{\emptyset}=1\)\footnote{We will now focus
on gauges where \(\Qs_{\emptyset}=1\), in this whole article,
since \(\Qs_{\emptyset} \ne 1\) case can be easily derived from the formulae of this paper by a gauge transformation.
}, and
  the whole set of non-zero  T-functions is given by the following three formulas

\begin{align}
\Ts_{1,s}&=+\Qs_{\ja}^{[s]}\Qs_{\overline{\ja}}^{[-s]}-
\nn \Qs_{\jb}^{[s]}\Qs_{\overline{\jb}}^{[-s]}\;\;,\;\;s\geq +1
\\
\Ts_{2,s}&=+\Qs_{\ja\jb}^{[s]}\Qs_{\overline{\ja\jb}}^{[-s]}\;\;,\;\;s\geq+2
\la{eq:explicitT} \\
\nn \Ts_{a,+2}&=
+\Qs_{\ja\jb}^{[a]}\Qs_{\overline{\ja\jb}}^{[-a]}\;\;,\;\;a\geq 2\\
\nn \Ts_{a,+1}&=(-1)^{a+1}\lb\Qs_{\ja\jb\jA}^{[a]}\Qs_{\overline{\ja\jb\jA}}^{[-a]}-
\Qs_{\ja\jb\jB}^{[a]}\Qs_{\overline{\ja\jb\jB}}^{[-a]}+
\Qs_{\ja\jb\jC}^{[a]}\Qs_{\overline{\ja\jb\jC}}^{[-a]}-
\Qs_{\ja\jb\jD}^{[a]}\Qs_{\overline{\ja\jb\jD}}^{[-a]}\rb\;\;,\;\;a\geq 1\\
\nn \Ts_{a,0}&=+\Qs_{\ja\jb\jA\jB}^{[a]}\Qs_{\jd\jc\jD\jC}^{[-a]}-
\Qs_{\ja\jb\jA\jC}^{[a]}\Qs_{\jd\jc\jD\jB}^{[-a]}+
\Qs_{\ja\jb\jA\jD}^{[a]}\Qs_{\jd\jc\jC\jB}^{[-a]}+
\Qs_{\ja\jb\jB\jC}^{[a]}\Qs_{\jd\jc\jD\jA}^{[-a]}-
\Qs_{\ja\jb\jB\jD}^{[a]}\Qs_{\jd\jc\jC\jA}^{[-a]}+
\Qs_{\ja\jb\jC\jD}^{[a]}\Qs_{\jd\jc\jB\jA}^{[-a]}\\
\nn \Ts_{a,-1}{}&=(-1)^{a+1}\lb\Qs_{\overline{\jd\jc\jD}}^{[a]}\Qs_{\jd\jc\jD}^{[-a]}
-
\Qs_{\overline{\jd\jc\jC}}^{[a]}\Qs_{\jd\jc\jC}^{[-a]}
+
\Qs_{\overline{\jd\jc\jB}}^{[a]}\Qs_{\jd\jc\jB}^{[-a]}
-\Qs_{\overline{\jd\jc\jA}}^{[a]}\Qs_{\jd\jc\jA}^{[-a]}
\rb\;\;,\;\;a\geq 1\\
\Ts_{a,-2}&=
\nn \Qs_{\overline{\jd\jc}}^{[a]}\Qs_{\jd\jc}^{[-a]}\;\;,\;\;a\geq 2\\
\nn \Ts_{2,s}&=+\Qs_{\overline{\jd\jc}}^{[-s]}\Qs_{\jd\jc}^{[s]}\;\;,\;\;s\leq-2\\
\nn \Ts_{1,s}&=+\Qs_{\overline{\jd}}^{[-s]}\Qs_{\jd}^{[s]}-\Qs_{\overline{\jc}}^{[-s]}\Qs_{\jc}^{[s]}\;\;,\;\;s\leq -1\\
\nn \Ts_{0,s}&=+\Qs_{\overline{\emptyset}}^{[-s]}\;
\end{align}
where the bar over the indices denotes the
complementary\footnote{The bar denotes the complementary
    set when \(I\)
  is sorted in increasing order for the canonical ordering
  \(\ja<\jb<\jc<\jd<\jA<\jB<\jC<\jD\). For instance, \(
  \Qs_{\overline{\ja\jb}}=\Qs_{\jc\jd\jA\jB\jC\jD}\).
 If the set \(I\) is not ordered this way, one has to commute its element first (which
 introduces a sign), and then to use this definition of the bar. For
 instance \(
  \Qs_{\overline{\jd\jc}}=-\Qs_{\overline{\jc\jd}}=-\Qs_{\ja\jb\jA\jB\jC\jD}\).
} set, for
instance \( \Qs_{\overline{\ja\jb}}=\Qs_{\jc\jd\jA\jB\jC\jD}\).\footnote{
Each term of these formulae should be T-functions related to infinite dimensional representations
of \(gl(4|4)\).
Let us introduce functions
\(\hat{\Ts}_{1,s} =\Qs_{1}^{[s]}\Qs_{\overline{1}}^{[-s]}- \Qs_{2}^{[s]}\Qs_{\overline{2}}^{[-s]}\) and
\(\check{\Ts}_{1,s}=-\Qs_{3}^{[s]} \Qs_{\overline{3}}^{[-s]} + \Qs_{4}^{[s]}\Qs_{\overline{4}}^{[-s]}\).
Note that the above \(\Ts^{\rm T-hook}_{1,s}\) coincides with \(\hat{\Ts}_{1,s}\) for  \(s \in {\mathbb Z}_{\ge 1}\)  and
 \(\check{\Ts}_{1,s}\) for \(s \in {\mathbb Z}_{\le -1}\) which correspond to infinite dimensional {\it irreducible} representations.
Then we find that the T-function related to finite dimensional irreducible \(s\)-th
symmetric tensor representation
 of \(gl(4|4)\)
is a subtraction of these functions:
\( \Ts^{\rm L-hook}_{1,s}=\hat{\Ts}_{1,s}-\check{\Ts}_{1,s}
\) \cite{Tsuboi:2009ud}.
Thus we see that an analytic continuation of
\(\check{\Ts}_{1,s}\) to positive \(s\) should be a T-function for an
 infinite dimensional {\it reducible} representations, which contains the finite dimensional irrep.
 This is similar to Bernstein-Gelfand-Gelfand-resolution on the level of the T-functions.
}
Note that this solution obeys the following properties:
\begin{equation}
(\p_u-2i\p_s)\Ts_{0,s}(u)=0,\qquad\Ts_{k,2}=\Ts_{2,k},\qquad \Ts_{k,-2}=\Ts_{2,-k},
\qquad  k \ge 2.
\end{equation} That means that we have imposed three out of four possible gauge constraints \eqref{eq:GaugeTr}.

Using the determinant formulas \eqref{detforQ5} and \eqref{detforQ6}
solving  the QQ-relations and the Laplace expansion formula for the determinants it is also possible
to rewrite T-functions in  various determinant forms.
For example, we have the following
 \(4\times 4\) and \(2\times 2\) determinant expressions reminding the character formula \eqref{CharSol}:
For the upper strip \(a\ge |s|\) of the \(\mathbb{T}\)-hook we can represent \(T_{a,s}\) as a \(4\times 4\) determinant
\begin{align}
\label{gl2M2M-3}
\Ts_{a,s}
&=\frac{(-1)^{sa}}
{{\cal (F}_{1-s}\circ\Qs_{\ja\jb}^{[a]}
{\cal )(F}_{1+s}\circ\Qs_{\jc\jd}^{[-a]})}
\begin{vmatrix}
(\Qs_{\ja\jb,j}^{[a-s-2i+3]})_{1\le i \le 2-s, \jA\le j \le \jD}  \\
(\Qs_{\jc\jd,j}^{[-a+s+3-2i]})_{1\le i \le 2+s, \jA\le j \le \jD}   \\
\end{vmatrix}.
\end{align}
and for the right strip (\(s\ge a \ge 0\))
\begin{align}
\label{gl2M2M-4}
\Ts_{a,s}
&=\frac{-1}
{{\cal F}_{1-a}\circ \Qs_{\jc\jd \jA\jB\jC\jD}^{[-s]} }
\begin{vmatrix}
\left(\Qs_{j,\jc\jd \jA\jB\jC\jD}^{[-s+a-3+2k]}\right)_{\substack{\ja \le j \le
    2 \\  1\le k \le 2-a}}  &
\left(\Qs_{j}^{[s-a+2k-1]}\right)_{\substack{\ja\le j \le 2 \\  1\le k \le a}}
  \\
\end{vmatrix}
\end{align}
and the left strip (\(-s\ge a \ge 0\)):
\begin{align}
\label{gl2M2M-5}
\Ts_{a,s}
&=\frac{-1}
{{\cal F}_{1-a}\circ \Qs_{\ja\jb\jA\jB\jC\jD}^{[-s]} }
\begin{vmatrix}
\left(\Qs_{j}^{[s-a+2k-1]}\right)_{\substack{\jc\le j \le \jd \\ 1\le k \le a}}  &
\left(\Qs_{\ja\jb,j,\jA\jB\jC\jD}^{[-s+a-3+2k]}\right)_{\substack{\jc\le j \le \jd\\ 1\le k \le 2-a}}   \\
\end{vmatrix}
\end{align}
as  \(2\times 2\) determinants.

Note that the above \(T\)-functions are taken in a different gauge than the \(T\)-functions of the generating functional. They are related as
\(\Ts^{[0]}_{a,s}=\Qs_{\emptyset}^{[s-a]} \Qs_{\ja\jb\jc\jd\jA\jB\jC\jD}^{[-s+a]} T^{[0]}_{a,s}\),
where \(\Qs_{\emptyset}=1\) in our normalization.
It is not difficult to see that these formulas solve indeed the Hirota equation in \(\mathbb{T}\)-hook.
Indeed, the Wronskian determinants \eqref{gl2M2M-3}, \eqref{gl2M2M-4} and \eqref{gl2M2M-5} have
similar structures
\footnote{
However one has to note that the matrix elements and
the prefactors of the Wronskians are quite non-trivially related each other
by the QQ-relations, in this case.}
 as the ones solving the Hirota equations in the infinite strips of a sizes \(N=4,2,2\), respectively (see
\cite{Krichever:1996qd}, and its applications \cite{Kazakov:2010kf}).

It is easy to see that to saw the three semi-infinite strips together into the full \(\mathbb{T}\)-hook it is enough to prove that the Wronskian solution satisfies four Hirota equations corresponding to the nodes \((a,s)=(\pm 1,\pm 2)\).
This  can be done straightforwardly.

There exist many possible determinant representations for the solution.\footnote{Some of them  will be published elsewhere \cite{FuturePaper2}.} In the next subsection we will
show, as an example, how to write every \(\Qs\)-function explicitly in terms of a basis of 8 \(\Qs\)
functions.

\subsection{A basis for $\Qs$ functions}

\label{sec:QBasis}

The aim of this subsection is to show that it is possible to express {\it explicitly and in a finite form} all the T-functions of the \(\mathbb{T}\)-hook in terms of a basis of 8 Q-functions, at least for a particular choice of this basis.

As we already mentioned, all the T-functions can be expressed, as the general solution of Hirota bilinear difference equation, through 8 independent functions.  Already the generating functional \eqref{eq:GenFn224} performs this task
 but, unfortunately, every T-function is expressed only by an infinite series in terms of 8 \(\xi\)-functions \eqref{eq:GF224notations},
organized in powers of wrapping.
Another solution involving infinite series was recently proposed in \cite{Hegedus:2009ky} which is closely related to the generating functional
\eq{eq:GenFn224} presented several years ago \cite{Beisert:2005di} in a different context.
Their parameterization \eqref{eq:chi-Q} in terms of 8 particular Q-functions does not immediately lead to finite expressions for the T-functions since to express the Q-functions of \eqref{eq:explicitT} in terms of these 8 Q-functions we need to solve the appropriate QQ-relations which, a priori, will lead again to the inversion of some finite difference operators, and hence to some infinite products.

Nevertheless, we will show here  that all the \(\Qs\) functions can be explicitly  expressed in a finite form as
Wronskian determinants in terms of
a particular basis \(\DbO\) where \(q_\jF=\Qs_{\overline{\jF}}/{\Qs_{\overline{\emptyset}}^+}\).
First,  the bosonization trick (see Appendix \ref{sec:RQQ}) says that by
relabelling all the \(\Qs\)-functions as \(\Qs_{I}=\tilde{\Qs}_{I\triangle B}\), defined by \eqref{eq:BosFermDef},
one gets \(\tilde \Qs\)-functions which obey for any indices (for the gradings with or without hats) only  the ``bosonic'' QQ-relation \eqref{eq:QQb} with
\(\ps_i=\ps_j=-1, \forall i,j\). In other words, the functions \(\tilde\Qs\)
don't distinguish bosonic and fermionic indices, and all the  \(\tilde\Qs\)-functions  with more than one index can be expressed through one-index    \(\tilde\Qs\)-functions
 by  standard bosonic determinants, namely
\begin{equation}
\label{eq:bosTrickDet}
  {\tilde \Qs}_{{i_1,i_2,\cdots
      i_N}}=\frac{\left|\left(\tilde\Qs_{{i_k}}^{[-N-1+2
        l]}\right)_{1\leq k,l\leq N}\right|}{{\cal F}_{N-1}\circ\tilde\Qs_{\emptyset}}
\end{equation}

In this manner, every \(\Qs\)-function  can be represented as a determinant of the
  Q-functions of the set
\(\mathcal{B}_2=\{{\tilde\Qs}_{\emptyset},{\tilde\Qs}_{\ja},{\tilde\Qs}_{\jb},{\tilde\Qs}_{\jc},{\tilde\Qs}_{\jd},{\tilde\Qs}_{\jA},{\tilde\Qs}_{\jB},{\tilde\Qs}_{\jC},{\tilde\Qs}_{\jD}\}\).

We will show now that we can express {\it\ all} \(\Qs\)-functions through 8 functions of the set
\(\mathcal{B}_1\) in an {\it\ explicit} and {\it\ finite} form.   On the other hand, \(\mathcal{B}_2\) contains \(9\)
\(\Qs\)-functions, which are not all
independent. Due to the Pl\"ucker relations among various
\(\Qs\)-functions given in the Appendix \ref{sec:RQQ},  these \(9\)
functions can be
expressed as determinants of  \(8\) independent functions of the  basis
\(\mathcal{B}_1\) as follows\footnote{In terms of the  gauge fixing,
  decreasing the number of independent functions from 9 to 8
  corresponds to adding the gauge constraint  \(\Qs_{\emptyset}=1\). It
  leaves another gauge degree of freedom yet unfixed :
    corresponding to a gauge transformation \(Q_I\to ({\cal
      F}_{n_b-n_f}\circ g)~Q_I\), where \(n_b\) and \(n_f\) are defined in
    footnote \ref{ft:nbf}. If we fix it, imposing for example \(\Qs_{\ja}=1\),  the number of independent functions becomes 7.}:
\begin{align}
\label{eq:B2B11}
  {\tilde\Qs}_{\emptyset}&=\left|{\left(\Qs_i^{[j]}\right)}_{\substack{\ja\leq
        i\leq \jd\\ j=3,1,-1,-3}}\right| \\
\label{eq:B2B12}
\tilde{\Qs}_{b}&=\left|\left({\Qs_i^{[j]}}\right)_{\substack{\ja\leq i\leq
      \jd;~i\neq b\\ j=2,0,-2}}\right|\,&
\ja\leq b\leq \jd\\
\label{eq:B2B13}
{\tilde \Qs}_{\jF}&=\Qs_{\overline{\emptyset}}^{[3]} \left|\left(q_\jI^{[j]}\right)_{\substack{\jA\leq \jI\leq
      \jD;~\jI\neq \jF\\ j=2,0,-2}}\right|&\jA\leq \jF\leq \jD
\end{align}
\begin{align}
\label{eq:B2B14}
\mathrm{where}\qquad\Qs_{\overline{\emptyset}}&={\tilde \Qs}_{\emptyset}^{[-4]} \Big/ \left|\left(
q_\jI^{[j]}
\right)_{\substack{\jA\leq \jI\leq
      \jD\\j=-1,-3,-5,-7}}\right| &q_\jF=\Qs_{\overline{\jF}}/{\Qs_{\overline{\emptyset}}^+}
\end{align}

The  relations \eqref{eq:B2B11} and \eqref{eq:B2B12}, for \(\tilde
\Qs_{\emptyset}\equiv\Qs_{\ja\jb\jc\jd}\) and  \((\tilde
\Qs_{\ja},\tilde \Qs_{\jb},\tilde \Qs_{\jc},\tilde
\Qs_{\jd})\equiv(\Qs_{\jb\jc\jd},\linebreak[1] \Qs_{\ja\jc\jd},\linebreak[1] \Qs_{\ja\jb\jd},\linebreak[1] \Qs_{\ja\jb\jc})\)
are  simply the determinant solution of bosonic Pl\"ucker QQ-relations
\eqref{detforQ5} (in the gauge \(\Qs_{\emptyset}=1\)).
In the same manner as the bosonization trick, the transformation
\(I\mapsto \bar I\) simply exchanges the gradings of all indices\footnote{Symbolically, we can still use \eqref{eq:QQb} and \eqref{eq:QQf}, together with the rule\\
\(
\ps_{\overline i}=-1,\quad i\in\{\ja,\jb,\jc,\jd\}\) and \(\ps_{\overline{\jI}}=1,\quad \jI\in\{\jA,\jB,\jC,\jD\}
\).
}: so that
the same bosonic determinant gives for instance
\beq
\Qs_{\overline{\jA\jB\jC}}= \frac{\det_{\substack{\jA\leq \jI\leq
      \jC\\j=2,0,-2}}\left({\Qs_{\overline{\jI}}^{[j]}}\right)}{
\Qs^+_{\overline{\emptyset}}\Qs^-_{\overline{\emptyset}}}\;\;,\;\; \text{and}\;\; \Qs_{\overline{\jA\jB\jC\jD}}=
\frac{\det_{\substack{\jA\leq \jI\leq
      \jD\\j=3,1,-1,-3}}
\left({\Qs_{\overline{\jI}}^{[j]}}\right)}{{\cal F}_3\circ
\Qs_{\overline{\emptyset}}}\;.
\eeq
 Due to the definitions \(({\tilde \Qs}_{\jA},{\tilde \Qs}_{\jB},{\tilde \Qs}_{\jC},{\tilde \Qs}_{\jD},{\tilde \Qs}_{\emptyset})=(\Qs_{\overline{\jB\jC\jD}},\Qs_{\overline{\jA\jC\jD}},\Qs_{\overline{\jA\jB\jD}}\Qs_{\overline{\jA\jB\jC}},\Qs_{\overline{\jA\jB\jC\jD}})\), these determinants are easily recast
into the formulae \eqref{eq:B2B13} and \eqref{eq:B2B14}.

In this way, we  fulfilled the task of this  subsection: all \(\Ts\)-functions inside the \(\mathbb{T}\)-hook can be expressed through 8  (ratios of) \(\Qs\)-functions.

The whole procedure is illustrated in the attached \textit{Mathematica}
  file, which for instance computes the expression of any
  \(\Qs\)-function in terms of the basis \(\mathcal{B}_1\).

\section{Asymptotic expressions for the generating functional and $\Qs$ functions}
\label{sec:ABA}

Now we will demonstrate our generating functional and Wronskian
solutions of the \linebreak[4] AdS/CFT Y-system in the asymptotic, large  size limit  \(L\to\infty\). We will present the asymptotic expressions of the relevant \(\chi\)- and \(\Qs\)-functions. All other \(\Qs\)-functions can be expressed through the basic ones through the Wronskian  relations described above.   Although it will be just a recasting of the ABA formulas of \cite{Beisert:2005tm} we feel that this Wronskian formulation is a right step in the direction of the derivation of the finite \(L\) system of FiNLIE's for the planar AdS/CFT spectrum.

\subsection{Asymptotic limit for the generating functional}
In the asymptotic limit the expressions for \(\xi\)-functions entering the generating functional \eqref{eq:Mon-chi},\eqref{eq:GenFn224} can be written explicitly in
terms of the ABA Bethe roots. Namely, expliciting the \(\Qs\)-functions in the general expressions \eqref{eq:chi-Q}
we write \cite{Gromov:2010vb,GromovKazakovReview} (see also \cite{Beisert:2006qh})
\begin{align}
\nonumber
&\xi_{\hat 1}\simeq H \mbar F^{+}\frac{B_3^-R_1^{-}}{B_3^+R_1^{+}}\;,\;\xi_1\simeq H\frac{B_3^-R_1^{-}Q_2^{++}}{B_3^+R_1^{+}Q_2 } \;\;,\;\;
\xi_2\simeq H\frac{B_1^+R_3^{+}Q_2^{--}}{B_1^-R_3^{-}Q_2}\;\;,\;\;
\xi_{\hat 2}\simeq H
\frac{B_1^+R_3^{+}}{B_1^-R_3^{-}}F^{-}
\\
&\xi_{\hat 4}\simeq \pbar H \frac{1}{\mbar F^{-}}\frac{B_5^+R_7^{+}}{B_5^-R_7^{-}}\;\;,\;\;
\xi_4\simeq \pbar H\frac{B_5^+R_7^{+} Q_6^{--}}{B_5^-R_7^{-} Q_6}\;\;,\;\;
\xi_3\simeq \pbar H
\frac{Q_6^{++} B_7^-R_5^-}{Q_6B_7^+R_5^{+}}\;\;,\;\;
\xi_{\hat 3}\simeq \pbar H\frac{B_7^-R_5^-}{B_7^+R_5^+ }\frac{1}{F^{+}}
\label{eq:chi-BRH}
\end{align}

where
\beq
F=\frac{R^{(-)}}{R^{(+)}}\;\;,\;\;
\mbar F=\frac{B^{(+)}}{B^{(-)}}
\eeq
and
\beq\label{eq:Hparam}
H=\lb\frac{x^{-}}{x^{+}}\rb^{\tfrac{L}{2}}\frac{B^{(+)+}}{B^{(+)-}}S(u)
\;\;,\;\; \pbar H=\lb\frac{x^{+}}{x^{-}}\rb^{\tfrac{L}{2}}\frac{B^{(-)-}}{B^{(-)+}}\frac{1}{S(u)}
\eeq
By \(\mbar{A}\) and \(\pbar A\)
  we denote the complex conjugations of \(A\) in the
  mirror and physical sheets, respectively (see the definition in \cite{Arutyunov:2007tc,Gromov:2009bc}).

Note that in the \(L\to\infty\) limit the first \(4\) \(\xi\)'s are suppressed in the mirror sheet whereas the last \(4\) are exponentially large.
The expansion of the generating series can be interpreted as an expansion in the powers of a small exponential  (which can be associated with  the first factor in the first of the formulas  \eqref{eq:Hparam}). It is then obvious that
this expansion, which can be also qualified as an expansion in ``wrappings'' associated with the SYM planar diagrams wrapping around the operator, can be also understood as a series  in powers of \(D=e^{-\frac{i}{2}\p_u}\) in the generating functional \eqref{eq:GenFn224}.
The quantities \(R_l^{(\pm)}\) and \(B_l^{(\pm)}\) are defined by
\begin{equation}
  R_l^{(\pm)}(u)\equiv \prod_{j=1}^{K_l}
  \left(
    {x(u)-x_{l,j}^{\mp}}
  \right)
  ,\quad
  B_l^{(\pm)}(u)\equiv\prod_{j=1}^{K_l}
  \left(
    \frac{1}{x(u)}-x_{l,j}^{\mp}
  \right),\qquad 1\leq l\leq 7
\end{equation}
where \(x_{l,j}^{\phantom{\pm}}\equiv x(u_{l,j}^{\phantom{\pm}}),\quad j\leq K_l\) encode the positions of Bethe roots. When the subscript \(l\)
isn't specified, the value \(l=4\) is assumed, and by definition \(x_{l,j}^{\pm}\equiv x(u_{l,j}^{\phantom{\pm}}\pm \frac{i}{2})\). The dressing
factor is
\(S(u)=\prod_j\sigma(x(u),x_{4,j}^{\phantom{\pm}})\) where \(\sigma\) is the BES dressing
kernel \cite{Beisert:2006ez}  (see \cite{Dorey:2007xn,Volin:2009uv} for a nice integral representation of the dressing kernel).

Furthermore, we define
\begin{equation}
\label{eq:defQl}
  Q_l^{(\pm)}=B_l^{(\pm)}
  R_l^{(\pm)}=\left(\prod_{j=1}^{K_l}-\frac{x_{l,j}^{\mp}}g\right)
  \prod_{j=1}^{K_l}(u^{\pm}-u_{l,j}^{\phantom{\pm}}),\qquad  1\leq l\leq 7
\end{equation}

\subsection{Explicit expressions  for the asymptotic $\Qs$-functions}

In the asymptotic (\(L\to \infty\)) limit, where we know the expressions of all
 monomials \(\xi\), all  \(\Qs\)
functions can be written in the leading ABA approximation  in terms of Wronskian determinants.

Matching the relevant ratios of  \(\Qs\)-functions from \eqref{eq:chi-Q}  with the expressions \eqref{eq:chi-BRH} we can express  the \(\Qs\) functions of the basis \(\DbO\)
of the previous section in the
asymptotic limit in the  explicit analytic form given in the
full generality in the appendix \ref{sec:AsQ}. We demonstrate it bellow
in a particular case of the \(SL(2)\)
sector, which consists of the states such that \(\forall l\neq 4,
K_l=0\),  implying that \(B_l, R_l\) and \(Q_l\) are equal to \(1\) when
\(l\neq 4\). The basis of 8 independent Q-functions looks as follows:
\begin{align}
\label{eq:B1sl2ab}
  \Qs_{\ja}&=h (1-\mbar F) & \Qs_{\jb}&=\Qs_{\ja}\cdot\left(-i u
      + \frac 1 2
  \frac{\mbar{F}+1}{\mbar{F}-1}
\right)\\
\label{eq:B1sl2cd}
\Qs_{\jc}&=\mbar F\cdot \frac {1}{h} (1-\frac 1 F) & \Qs_{\jd}&=\Qs_{\jc}\cdot\left(-i u
      + \frac 1 2
  \frac{{F}+1}{{F}-1}
\right)\\
\label{eq:B1sl2AB}
q_\jA&=
\left(-M-\frac{Q_{}^{-}}{2 F}+\frac{F~Q_{}^{+}}{2}\right) \cdot q_{\jB}&
q_\jB&= f^+ h \mbar{F}\\
\label{eq:B1sl2CD}
q_\jC&=
\left(-M-\frac{F~Q_{}^{+}}{2}+\frac{Q_{}^{-}}{2 F} \right) q_{\jD}&
q_\jD&=\frac{\mbar{f}^-}{\pbar{h}}
\end{align}

where \(f\), \(\mbar{f}\), \(h\) and \(\pbar h\) are defined by the recursion relations
\begin{equation}
  \frac {h^+}{h^-}=H,\qquad \frac {\pbar h^-}{\pbar h^+}=\pbar
  H,\qquad \frac {f^+}{f^-}=F,\qquad \frac {\mbar{f}^-}{\mbar{f}^+}=\mbar{F}
\end{equation}
while \(M\) is a polynomial function of \(u\), defined by
the recursion relation
\begin{eqnarray}
\label{eq:DefM}
  M^+-M^-&=&2 Q-Q\left(F^-+\frac 1 {\mbar F ^-}\right)
  -Q^{++}\left(F^++\frac 1 {\mbar F ^+} \right)\,.
\end{eqnarray}
In the RHS of \eqref{eq:DefM} we find a sum of two terms both having the form \(Q^+ \left(F+\frac
    {1} {\mbar {F}
    }\right)=\frac{Q^{+}}{Q^{(+)}}\left(R^{(-)}B^{(+)}+B^{(-)}R^{(+)}\right)\)
  but with different shifts of argument. It can be easily  seen
    that they are  polynomials
\footnote{\label{ft:polyn}First, it is clear
    from \eqref{eq:defQl} that \(\frac{Q^{+}}{Q^{(+)}}
=
\prod_{j=1}^{K_4} \frac{x_{4,j}}{x_{4,j}^-}
\)
 is a
    constant. Then one can expand
\(B^{(+)}R^{(-)}+B^{(-)}R^{(+)}=\left(\prod_j
    {x(u)-x_{4,j}^{-}}
  \right)\left(\prod_j
    {\frac 1 {x(u)}-x_{4,j}^{+}}
  \right)+\left(\prod_j
    {x(u)-x_{4,j}^{+}}
  \right)\left(\prod_j
    {\frac 1 {x(u)}-x_{4,j}^{-}}
  \right)\) : the first product gives terms of the form
  \(\left(x_{4,i_1}^+,x_{4,i_2}^+,\cdots,x_{4,i_{m}}^+\right)
  \left(x_{4,j_1}^-,\cdots,x_{4,j_{n}}^-\right) x(u)^{n-m}\),
  while the second product gives terms of the form
\(\left(x_{4,i_1}^-,x_{4,i_2}^-,\cdots,x_{4,i_{m}}^-\right)
  \left(x_{4,j_1}^+,\cdots,x_{4,j_{n}}^+\right)
  x(u)^{n-m}\). Then the definition of the Zhukovsky map
  \(\frac{u}{g}=x+1/x\) allows to rewrite each term
  \(\left(x_{4,i_1}^+,x_{4,i_2}^+,\cdots,x_{4,i_{m}}^+\right)
  \left(x_{4,j_1}^-,\cdots,x_{4,j_{n}}^-\right)
  \left(x(u)^{n-m}+x(u)^{m-n}\right)\) as a polynomial in \(u\).
}
 so that \(M\) itself is a
    polynomial which can be found (up to an additive constant) from the eq.\eqref{eq:DefM} by matching
    the coefficients of the RHS and the LHS.

For instance, for Konishi state, there are two roots
\((u_{4,1},u_{4,2})\equiv (\mathfrak u_1, -\mathfrak u_1)\), so that
\({\varkappa}_2^{\pm}\equiv x_{4,2}^{\pm}=-x_{4,1}^{\mp}\equiv -{\varkappa}_1^{\mp}\). Then\footnote{In these expressions,
  \(\varkappa_1^\pm\) denotes \(x(u_{4,1}^{\phantom{\pm}}\pm \frac{i}{2})\).} \(Q^+
\left(F+\frac
    {1} {\mbar {F}
    }\right) = 2~{\varkappa_1^{}}^2\left(4-\frac {u^2}{g^2}+\left(\varkappa_1^+-\frac 1
      {\varkappa_1^+}\right) \left(\varkappa_1^--\frac 1 {\varkappa_1^-}\right)  \right)
  \), and \eqref{eq:DefM} is solved by
\(
M~ =~ 2~ i~ u~{\varkappa_1^{}}^2\left(
-\frac{{\mathfrak u _1^{}}^{2}}{g^2}-\frac 1 3 \frac{u^+ u^- -1}{g^2}
+ 8 + 2\left(\varkappa_1^+-\frac 1
    {\varkappa_1^+}\right) \left(\varkappa_1^--\frac 1 {\varkappa_1^-}\right)\right)
\), which is indeed a polynomial in \(u\).

\subsection{Physical symmetries}
Some symmetries can be identified in this asymptotic solution, which
can be viewed as generalizations of  symmetries of the characters
of the classical monodromy matrix.
\paragraph{Lower boundary:}
\label{sec:lower}
The first observation is that \(\Qs_{\overline \emptyset}=1\). This could already
be immediately seen
from
  \eqref{eq:chi-BRH} by computing \(\frac{\Qs_{\overline
  \emptyset}^+}{\Qs_{\overline \emptyset}^-}= {\left(\frac
  {\xi_{\jd}\xi_{\jb}} {\xi_{\jD}\xi_{\jB}} \right)^+ \left(\frac
  {\xi_{\jc}\xi_{\ja}} {\xi_{\jC}\xi_{\jA}} \right)^-}=1 \). This is a generalization of the fact that the classical monodromy
matrix \(\Omega(x)\in SU(2,2|4)\) has the super-determinant equal to \(1\), and in
terms of \(\Ts\) functions, it means, by virtue of the last relation from \eqref{eq:explicitT}, that \(\Ts_{0,s}=1\) (as explained in Appendix \ref{sec:Hda} we can drop a possible \(i\)-periodic factor in \(\Qs_{\overline{\emptyset}}\)).

\paragraph{Mirror reality:}
The reality  of the \(\Ts\) functions on the mirror sheet, which is in the
asymptotic limit a consequence of the relations \cite{GromovKazakovReview}
\(\xi_{\ja}=\mbar{\xi_{\jb}}\), \(\xi_{\jA}=\mbar{\xi_{\jB}}\),
\(\xi_{\jc}=\mbar{\xi_{\jd}}\),
\(\xi_{\jC}=\mbar{\xi_{\jD}}\), then translates into the
condition that
\footnote{We use the definition \((-1)^{\jI}\equiv(-1)^{i}\).}
 \begin{equation}
\label{eq:ReQ}
  \Qs_{\overline{I}}=\left(\prod_{i\in I}(-1)^{i}\right)\left({\cal F}_{n_b-n_f} \circ g_{1}\right)\mbar{\Qs_{\mathrm{Ex}(I)}}
\end{equation}
which involves the gauge function \(g_1=\frac 1 {F \mbar{F}}\),
while \(n_f\) and
\(n_b\) are the number of ``bosonic'' and ``fermionic'' indices\footnote{\label{ft:nbf}\(n_b=\mathrm{Card}I\cap \{\ja,\jb,\jc,\jd\}\) and
  \(n_f=\mathrm{Card}I\cap \{\jA,\jB,\jC,\jD\}\)} in
\(I\), and the index
exchange function
\(I\mapsto \mathrm{Ex}(I)\), transforms all indices in \(I\)
as follows: \(\ja\leftrightarrow\jb,\jc\leftrightarrow\jd,\jA\leftrightarrow\jB,\jC\leftrightarrow\jD\).

For instance, when \(I=\{\jc,\jd,\jB,\jC,\jD\}\), the equation \eqref{eq:ReQ}
states that\footnote{
In \eqref{eq:Reg}, the second equality is given by \eqref{eq:ReQ},
  and the sign \(+\) comes from \(\prod_{i\in
    I}(-1)^{i}=(-)(+)(+)(-)(+)=+1\). The third equality reorders the
  set of indices, and involves the sign of a permutation. This
  permutation exchanges the positions of two pairs of indices, hence the
\(+\) sign in the third equality.
}
\begin{equation}
\label{eq:Reg}
\Qs_{\ja\jb\jA}=\Qs_{\overline{\jc\jd\jB\jC\jD}}= +~\frac 1 {g_1} ~
\mbar{\Qs_{\jd\jc\jA\jD\jC}}= +~\frac 1 { g_1 }~  \mbar{\Qs_{\jc\jd\jA\jC\jD}}
\end{equation}

\paragraph{Wing exchange transformation}
Another symmetry of the asymptotic solution corresponds to the fact  that, up to a gauge, \(\Ts_{a,-s}\) is
obtained from \(\Ts_{a,s}\) by the transformation
\(\mathcal{WE}:\,\,\{B_{k}\leftrightarrow B_{8-k},\,\,\,
R_k\leftrightarrow R_{8-k}\}\).
In terms of \(\Qs\) functions, this relation reads
\begin{equation}
\label{eq:WE}
  \forall I,  \qquad \Qs_I=(-1)^{n_b}(-1)^{\left\lfloor \frac {n_b+n_f} 2\right\rfloor}\left({\cal F}_{n_b-n_f} \circ
     g_2\right)\mathcal{WE}\left( \Qs_{\overline{\mathrm{WE}(I)}}\right)
\end{equation}
where the Wing
Exchange function \(I\mapsto \mathrm{WE}(I)\), transforms the individual indices in  \(I\) according to \(
b\leftrightarrow \je-b,\quad
\jF\leftrightarrow \jE-\jF
\equiv \widehat{\je-f} \), where \(\ja\leq b\leq \jd\) and \(\jA\leq \jF\leq \jD
\) and \(x\mapsto \lfloor x\rfloor\) denotes the floor function.
For instance, for \(I=\{\ja,\jb,\jA\}\), \eqref{eq:WE} states that\footnote{
The first equality is given by \eqref{eq:WE},
  and the sign \(-\) comes from \((-1)^{2+\left\lfloor \frac {3}
      2\right\rfloor}=-1\). The second equality reorders the
  set of indices, and involves the sign of a permutation.}
 \(
  \Qs_{\ja\jb\jA}=-g_2 \mathcal{WE}\left(\Qs_{\overline{\jd\jc\jD}}\right)=g_2 \mathcal{WE}\left(\Qs_{\ja\jb\jA\jB\jC}\right)
\),
 which can be written as \(h \mbar{f}^- B_1 R_3 = g_2 \mathcal{WE}
\left( \pbar{h} \mbar{f}^+ B_7 R_5 \right)
\), hence the value of the  gauge function implied in that relation is \(g_2=\mbar{F}^2\).

\subsection{Comments to the asymptotic Wronskian solution}

These equations simply recast \eqref{eq:GenFn224} into the form
  of determinants involving only the functions of the basis
  \(\mathcal{B}_1\). The size of these determinants is fixed and doesn't
increase with \((a,s)\).

On the other hand, the asymptotic \(\Qs\)-functions of \(\mathcal{B}_1\) involve the
functions \(f\) and \(h\), which are expressed through infinite products
and cannot be avoided in \(\Qs\) functions, because we have to express them knowing explicitly  (in terms of \(\xi\) functions) only the \(\Qs_I^+/\Qs_I^-\). But fortunately in the \(\Ts\) functions these infinite products are absent since  the \(T\)-functions are made of products of the type \(\Qs_I^{[+s]}
\Qs_{\overline{I}}^{[-s]}\) involving only finite products
\(\frac{h^{[+s]}}{h^{[-s]}}={\cal F}_s\circ H\).

Similarly, although the \(M_{i,j}\) functions look like infinite
  sums\footnote{In is not the case  for instance in the \(sl(2)\) sector,
    because the \(M\) functions are
    polynomial and are easily identified, and  no  infinite
  sum arises.} it is possible to show that \(T\)-functions involve only differences of the form
\(M_{i,j}^{[+s]}-M_{i,j}^{[-s]}\), which are finite sums.
 One can see it by  certain subtractions
of  columns or lines of the determinants (\ref{gl2M2M-3}-\ref{gl2M2M-5}) which does not change their values.

In conclusion,   we demonstrated in this section our general solution of Hirota equation given in the previous section,  in the case of asymptotic, large \(L\) limit of the AdS/CFT Y-system.  As our experience with the principal chiral field model showed  \cite{Gromov:2008gj,Kazakov:2010kf} such asymptotic Wronskian expressions in terms of well chosen Q-functions can be very useful in establishing the finite \(L\) solution of the sigma model under investigation.

\section{Conclusion}

The main purpose of this paper was to express all the Y-functions and the associated T-functions entering the  \(\mathbb{T}\)-hook of the AdS/CFT Y-system, using its  discrete  integrable Hirota dynamics, in the form of an explicit Wronskian
determinant expressions parameterized through a finite set   of  8  Baxter-type Q-functions (7 independent Q-functions after all the gauge constraints are imposed).
We view it as an important step towards the derivation of a Destri-deVega-like {\it finite} system of non-linear integral equations  (FiNLIE)  for this important model.

For certain relativistic sigma-models these Wronskian expressions,  due to the integrable discrete Hirota dynamics of the  Y-system,   gave us a direct access to the corresponding FiNLIE \cite{Gromov:2008gj,Kazakov:2010kf}.
In the case of  \(SU(N)\) principle chiral field the corresponding Q-functions were simple polynomials in the asymptotic large  \(L\) limit,
 and could be easily generalized to describe the  finite size \(L\) system  by  introduction of certain \(N-1\)
discontinuities, or ``densities'', vanishing at large \(L\), along the whole real axis of the spectral parameter \(u\). It suffices then to substitute these expressions into \(N-1\) TBA equations for the  momentum-carrying nodes to write the needed FiNLIE \cite{Kazakov:2010kf}.

We don't know yet how to introduce these densities in the AdS/CFT case although it is conceivable that they
could be non-zero only on the Zhukovsky-type cuts with the branch points at \(u=\pm 2g+\frac{in}{2}\) for some  \(n\in \mathbb{Z}\).
Apart from the analyticity in the spectral parameter \(u\), we also have to understand and incorporate into the Wronskian solution
the asymptotic properties of T-functions w.r.t. large values of \(a\) and \(|s|\).  The hope is that, as in the case of \(SU(N)\) principal chiral field,
there exists for the Wronskian solution of \(\mathbb{T}\)-hook the
``best'' basis of \(7\) Q-functions with the simplest possible analytic properties.
The analyticity of the rest of the quantities, Q-,T- or Y-functions, will be simply a consequence of
the analytic properties of these \(7\) Q-functions and of  the Wronskian formulas presented in this paper.
We hope to describe it in the future work.

Another interesting problem is to understand whether  our Wronskian expressions could  be promoted to an operatorial form.
In the quantum spin chains or even in the conformal field theories the Q-operators enter the same commuting family of operators as the T-operators (transfer matrices), and  operatorial form of Wronskian expressions
makes a perfect sense and can be constructed,
for example using the approach of 
\cite{Bazhanov:1996dr,Bazhanov:1998dq,Bazhanov:2001xm,Bazhanov:2008yc} using \(q\)-oscillator representations of the
quantum affine algebra and the universal R-matrix.

Let us also mention an interesting problem of the generalization of our Wronskian representation to another integrable duality,    AdS\(_4\)/CFT\(_3\), relating the 3D ABJM gauge theory and the sigma model on \(AdS_4\times CP^3\). The Y-system for this model is a known
 \cite{Gromov:2009tv,Bombardelli:2009xz,Gromov:2009at} solution and the corresponding Q-system was obtained for some cases in \cite{Gromov:2009at}.
 The Y-system contains only one wing and it could be easier to study than the  AdS\(_5\)/CFT\(_4\) duality.

 Apart from these main, physical tasks there are also a few other, more technical or mathematical  questions to understand in our formalism. There should exist a natural generalization of the Wronskian solution in the \((2|4|2) \)    \(\mathbb{T}\)-hook given
 here,    to all   \((M_1|N|M_2) \) \(\mathbb{T}\)-hooks related to the infinite dimensional representations of \(u(M_1,M_2|N)\) \cite{FuturePaper2}.
The method of the
``co-derivative'' \cite{KV07,KLTtoappear} looks especially promising and could also be useful for
 operator construction mentioned above.
This could potentially open an interesting field of research related to the search and investigation of a possible large
new class of non-compact sigma models.

The Wronskian determinant representation of the T-functions, solving Hirota equation in the AdS/CFT
related \(\mathbb{T}\)-hook in terms of a finite set of Q-functions, presented in this paper, gives
us reasonable hopes for the construction of    an AdS/CFT FiNLIE system, the analogue of Destri-deVega
equations,  and  for a deeper understanding of the physical nature of AdS/CFT integrability.

\section*{Acknowledgments}
The work  of NG and  VK was partly supported by  the grant RFFI 08-02-00287. The work  of VK was also partly supported by  the ANR grant   GranMA (BLAN-08-1-313695).  V.K. also thanks NORDITA institute in Stockholm, as well as  Vladimir Bazhanov and  the theoretical physics group of  Australian National University (Canberra) for the kind hospitality and interesting discussions on  this project.  We  thank
 Pedro Vieira and  Dmytro Volin for useful comments and discussions.
The work of ZT was
supported by Nishina Memorial Foundation and by
Grant-in-Aid for Young Scientists, B \#19740244 from
The Ministry of Education, Culture, Sports, Science and Technology in Japan.
ZT thanks Ecole Normale Superieure, LPT, where a considerable part of this work was done,
for the kind hospitality.
ZT also thanks Rouven Frassek, Tomasz Lukowski, Carlo Meneghelli and Matthias Staudacher at
Humboldt-Universit\"{a}t zu Berlin, Institut f\"{u}r Mathematik
 for the kind hospitality and discussions.
\appendix
\section{Relations between $Q$ functions}
\label{sec:RQQ}

These QQ-relations represent special versions of the general Pl\"ucker relations for determinants. That means that there exists a certain number of QQ-relations expressing some Q-functions through the other, leaving only \(8\) independent functions.  A few useful  determinant relations express Q-functions of a later level of nesting through the ones on an earlier stages, such as
\begin{multline}
\Qs_{I,b_{1} b_{2} \dots b_{m} f_{1} f_{2} \dots f_{n}}^{ }=
\frac{(-1)^{\frac{m(m-1)}{2}}
\begin{vmatrix}
(\Qs_{I,b_{j} f_{k}}^{[-m+n]})_{1 \le j \le m,1 \le k \le n} &
(\Qs_{I,b_{j}}^{[-m+n+2k-1]})_{1 \le j \le m,1 \le k \le m-n}
\end{vmatrix}
}
{\left(\Qs_{I}^{[-m+n]}\right)^{n} {\cal F}_{m-n-1}\circ \Qs_{I}}
\\ \text{for}\quad m \ge n  \label{detforQ5}
\end{multline}
and
\begin{align}
\Qs_{I,b_{1}b_{2}\dots b_{m}f_{1}f_{2}\dots f_{n}}^{ }
=
\frac{(-1)^{\frac{n(n-1)}{2}+m(n-1)}
\begin{vmatrix}
(\Qs_{I,b_{j} f_{k}}^{[-m+n]})_{1 \le j \le m,1 \le k \le n} \\
(\Qs_{I,f_{k}}^{[-m+n-2j+1]})_{1 \le j \le n-m,1 \le k \le n}
\end{vmatrix}
}
{\left(\Qs_{I}^{[-m+n]}\right)^{m} {\cal F}_{n-m-1}\circ \Qs_{I}}
\quad \text{for}\quad m \le n. \label{detforQ6}
\end{align}
For \(I=\emptyset \), these formulae reduce
\footnote{
In this sense, these formulae are
a generalization of determinant formulae of Theorem 3.2 in \cite{Tsuboi:2009ud}.
However, one can easily seen that these follow from Theorem 3.2
just by manipulating the index set of \(Q\)-functions.
For two tuples, \(I\) and \(J=(b_{1}b_{2}\dots b_{m}f_{1}f_{2}\dots f_{n})\) (\(I\) is fixed),
let us consider a gauge transformation
\begin{align}
\tilde{\Qs}_{I,J}^{ }=\frac{\Qs_{I,J}^{ }}{\Qs_{I}^{[\sum_{j \in J}\ps_{j}]}},
\end{align}
where `\(I,J\)' is a concatenation of
\(I\) and \(J\).
Since there exists a relation
\(\tilde{\Qs}_{I,\emptyset}^{ }=\tilde{\Qs}_{I}^{ }=1\), one can apply the
Theorem 3.2 for \(\tilde{\Qs}_{I,J}^{ }\).
Then we obtain the formulae \eqref{detforQ5} and \eqref{detforQ6}.
}
to the determinant solutions of the
\(QQ\)-relations in \cite{Tsuboi:2009ud}.
Next, we introduce
a useful trick on the index set for the determinant formulae,
which may be called ``bosonization'' or ``fermionization'' trick. Let us denote
\begin{align}\label{eq:BosFermDef}
\tilde{\Qs}_I=\Qs_{I\triangle B}, \quad
\check{\Qs}_I=\Qs_{I\triangle F}
\end{align}
where \(A\triangle C=(A \cup
C)\setminus(A \cap C)\),
\(B=(\ja,\jb,\jc,\jd)\) and \(F=(\jA,\jB,\jC,\jD)\). In other words, we define a \(\tilde{\Qs}_I\) through the corresponding \(\Qs_J\) by adding to its indices forming the set \(J\) all the indices from the set  \(B=(\ja,\jb,\jc,\jd)\) and then removing those of them which already were contained in  \(J\), to get finally \(I\).
\\ From \eqref{detforQ5}, we obtain
\begin{align}
\check{\Qs}_{I,c_{1} c_{2} \dots c_{m}}^{ }=
\frac{(-1)^{\frac{m(m-1)}{2}}
\begin{vmatrix}
(\check{\Qs}_{I,c_{j}}^{[-m+2k-1]})_{1 \le j,k \le m}
\end{vmatrix}
}
{ {\cal F}_{m-1}\circ \check{\Qs}_{I}}.  \label{detforQ7}
\end{align}
From \eqref{detforQ6}, we obtain
\begin{align}
\tilde{\Qs}_{I,c_{1}c_{2}\dots c_{n}}^{ }
=
\frac{(-1)^{\frac{n(n-1)}{2}}
\begin{vmatrix}
(\tilde{\Qs}_{I,c_{k}}^{[n-2j+1]})_{1 \le j, k \le n}
\end{vmatrix}
}
{{\cal F}_{n-1}\circ \tilde{\Qs}_{I}}.
\label{detforQ8}
\end{align}

This trick is efficient in the sense that it allows to write quite easily all \(\Qs\) functions in terms of only \(8\) functions, typically the single-indexed \(\hat {\Qs}\) functions.

\section{Asymptotic expression of $Q$ function}

\label{sec:AsQ}

In the general case (outside the \(sl(2)\) sector), the formulae
\eqref{eq:B1sl2ab}-\eqref{eq:B1sl2CD} become, in full generality :

\begin{eqnarray}
\label{eq:B1a}
    \Qs_{\ja}&=&\frac{h^{} \left(-\mbar{F}^{}
        Q_2^{-}+Q_2^{+}\right)}{B_3^{} R_1^{}}\\
\label{eq:B1b}
\Qs_{\jb}&=&\Qs_{\ja}\cdot\left(M_{\ja\jb} + \frac{Q_1^{}Q_3^{}}{ 2 Q_2^{+}Q_2^{-}}
  \frac{\mbar{F}Q_2^-+Q_2^+}{\mbar{F}Q_2^--Q_2^+}
\right)\\
\Qs_{\jc}&=&
\frac{\mbar{F}^{}\left(-Q_6^{-}+F^{} Q_6^{+}\right)}{F^{} h^{}
  B_7^{} R_5^{}}\\
\Qs_{\jd}&=&\Qs_{\jc}\cdot\left(-M_{\jc\jd}^{}+\frac{Q_5^{} Q_7^{}}{2 Q_6^{+} Q_6^{-}}
\frac{Q_6^-+F Q_6^+}{Q_6^--F Q_6^+}\right)\\
q_\jA&=&
\left(-M_{\overline{\jA\jB}}-\frac{\frac{Q_{}^{-}}{F}-F~Q_{}^{+}}{2 Q_1
    Q_3}\right) \cdot q_{\jB}\\
q_\jB&=& f^+ h \mbar{F} B_1 R_3\\
q_\jC&=&
\left(-M_{\overline{\jC\jD}}-\frac{F~Q_{}^{+}-\frac{Q_{}^{-}}{F}}{2 Q_5 Q_7} \right) q_{\jD}\\
q_\jD&=&\frac{\mbar{f}^- B_5 R_7}{\pbar{h}}
\label{eq:B1D}
\end{eqnarray}
where \(M_{\ja\jb}, M_{\jc\jd}, M_{\overline{\jA\jB}}\) and
\(M_{\overline{\jC\jD}}\) are meromorphic
functions (without cuts, as explained in Appendix \ref{sec:Hda}) on the complex plane defined by
the recursion relation
\begin{eqnarray}
  M_{\ja\jb}^+-M_{\ja\jb}^-&=&\frac 1 2 \left(\left(\frac{Q_1^{}
  Q_3^{}}{Q_2^{+}
      Q_2^{-}}\right)^++\left(\frac{Q_1^{}
  Q_3^{}}{Q_2^{+}
      Q_2^{-}}\right)^-\right)\\
M_{\jc\jd}^+-M_{\jc\jd}^-&=&\frac 1 2 \left(\left(\frac{Q_5^{}
  Q_7^{}}{Q_6^{+}
      Q_6^{-}}\right)^++\left(\frac{Q_5^{}
  Q_7^{}}{Q_6^{+}
      Q_6^{-}}\right)^-\right)\\
  M_{\overline{\jA\jB}}^+-M_{\overline{\jA\jB}}^-&=&
  \frac{-Q_{}^{}~F^--\frac{Q_{}^{--}}{F^-}+2 Q_{}^{}  \frac{Q_2^{--}}{Q_2}}
  {2 Q_1^-Q_3^-}
+  \frac{-\frac {Q_{}^{}}{F^+}-{Q_{}^{++}}{F^+}+2 Q_{}^{}  \frac{Q_2^{++}}{Q_2}}
  {2 Q_1^+Q_3^+}\\
  M_{\overline{\jC\jD}}^+-M_{\overline{\jC\jD}}^-&=&
  \frac{-Q_{}^{}~F^--\frac{Q_{}^{--}}{F^-}+2 Q_{}^{}  \frac{Q_6^{--}}{Q_6}}
  {2 Q_5^-Q_7^-}
+  \frac{-\frac {Q_{}^{}}{F^+}-{Q_{}^{++}}{F^+}+2 Q_{}^{}  \frac{Q_6^{++}}{Q_6}}
  {2 Q_5^+Q_7^+}\,.
\end{eqnarray}

Note that all these recursion relations could be solved analytically
in terms of infinite products, or of an integral representation taking
into account their analyticity properties.  In the \(sl(2)\) sector of quantum states of the AdS/CFT system, they can be
 identified  simply as polynomials.

We can check, by the direct substitutions of the formulas
\eqref{eq:B1a}-\eqref{eq:B1D} into the expressions \eqref{eq:chi-Q},
that they reproduce \eqref{eq:chi-BRH}, and hence the right asymptotic solution of the Y-system given in \cite{Gromov:2009tv}.

\subsection{Hints of derivation of these asymptotic $\Qs$ functions}
\label{sec:Hda}

Let us explain the derivation of \eqref{eq:B1a}-\eqref{eq:B1D}, starting from
the expressions of the \(\xi\)-factors \eqref{eq:chi-BRH}. We will see that
the derivation
 assumes that a certain reality condition is satisfied. It is explained here only for the
asymptotic limit  (because \eqref{eq:chi-BRH} gives an explicit expression to start with),
 but in fact it applies to the construction of
any real solution of the Y-system\footnote{
In the sense that \eqref{eq:ReQ} is satisfied for a gauge \(g_1\) which is a priori unknown.}, in terms of 7 independent Q-functions.

The first step is to rewrite \eqref{eq:chi-BRH} in terms of \(\Qs\)
function (by matching \eqref{eq:chi-BRH} with \eqref{eq:chi-Q}). This
gives
\(\Qs_{\jA}=\frac{\mbar f ^+ B_3
  R_1}{h}\), \(\Qs_{\ja\jA}=-\Qs_{\jA\ja}={Q_2\mbar f}\),
\(\Qs_{\ja\jb\jA}=\Qs_{\jA\ja\jb}={h \mbar f^- B_1 R_3}\),
\(\Qs_{\ja\jb\jA\jB}=\Qs_{\jA\ja\jb\jB}={\frac {\mbar f} f}\),
\(\Qs_{\ja\jb\jA\jB\jC}=\Qs_{\jA\ja\jb\jB\jC}={\pbar h B_7 R_5} \mbar f^+\), 
\(\Qs_{\ja\jb\jc\jA\jB\jC}=-\Qs_{\jA\ja\jb\jB\jC\jc}=\mbar f Q_6\),
\(\Qs_{\ja\jb\jc\jd\jA\jB\jC}=\Qs_{\jA\ja\jb\jB\jC\jc\jd}=\frac{\mbar
  f^- B_5 R_7}{\pbar h}\), and 
\(\Qs_{\overline{\emptyset}}=\Qs_{\jA\ja\jb\jB\jC\jc\jd\jD}=1\).

In particular it gives
\(\Qs_{\overline{\emptyset}}=1\).
 This parameterization is done up
to \(8\) arbitrary \(i\)-periodic
  functions \(v_j(u)=v_j(u+i),\,\,j\in I_0\):
the \(\xi\) functions do not change under the transformation \(\Qs_{I}\mapsto
  \prod_{i\in I} v_i^{[\mathrm{Card(}I)]} \Qs_I\) (where
  \(\mathrm{Card}(I)\) denotes the
  length of \(I\)), which leaves the QQ-relations unchanged, and
  leads to an \(i\)-periodic gauge transformation of the \(\Ts\)-functions.

Then the QQ-relation \eqref{eq:QQf} is used to find
\(\Qs_{\ja}=(\Qs_{\ja\jA}^+- \Qs_{\ja\jA}^-)/\Qs_{\jA} = \frac{h}{B_3
R_1}\left(Q_2^+-\mbar F Q_2^-\right)\). The same QQ-relation also gives\\
\beq\Qs_{\ja\jb}=\frac{\Qs_{\ja\jb\jA}^+ \Qs_{\ja}^- -
\Qs_{\ja\jb\jA}^- \Qs_{\ja}^+}{\Qs_{\ja\jA}}=\frac{h^+ h^-\mbar
  F^-}{Q_2}\left(\frac{B_1^-R_3^-}{B_3^+R_1^+}(\mbar F^+
  Q_2-Q_2^{++})+\frac{
  B_1^+R_3^+}{B_3^-R_1^-}(\frac{Q_2}{\mbar F^-}-Q_2^{--})\right)\eeq
Then the ``bosonic'' QQ-relation \eqref{eq:QQf} allows to write
\(\left(\frac{\Qs_{\jb}}{\Qs_{\ja}}\right)^+-\left(\frac{\Qs_{\jb}}{\Qs_{\ja}}\right)^-
= -\frac{\Qs_{\ja\jb}}{\Qs_{\ja}^+\Qs_{\ja}^-}\). By
plugging here the previous two expressions we get\footnote{with the
  same definition of \(\pbar A\) as the complex conjugate of \(A\) on physical sheet. The following formulas exhibit the physical reality property of all Q-functions in the asymptotic limit inherited from the \(Z_4\) symmetry of the classical finite gap solution.}
\begin{equation}
  \left(\frac{\Qs_{\jb}}{\Qs_{\ja}}\right)^+ -
  \left(\frac{\Qs_{\jb}}{\Qs_{\ja}}\right)^- =A^+-\pbar  A ^-\qquad
  A=\frac{Q_1 Q_3}{\mbar F^{} (Q_2^-)^2-Q_2^-Q_2^{+}}\qquad \pbar
  A=\frac{Q_1 Q_3}{\frac{(Q_2^+)^2}{\mbar F}-Q_2^+Q_2^{-}}
\end{equation}
Which is solved as
\begin{equation}
  \frac{\Qs_{\jb}}{\Qs_{\ja}}=M_{\ja\jb}+\frac 1 2 \left(A-\pbar
    A\right)\qquad \qquad  M_{\ja\jb}^+-M_{\ja\jb}^-= \frac 1 2
    (A^++\pbar A^++A^-+\pbar A^-)
\end{equation}

These two first steps give \eqref{eq:B1a} and \eqref{eq:B1b}. The same
manipulations help to find  \(\Qs_{\overline \jD}\) and \(\Qs_{\overline \jC}\) :
First, \(\Qs_{\overline \jD}\) is extracted directly from the \(\xi\)-functions,
and then the ``fermionic'' QQ-relations give successively
\(\Qs_{\overline{\jd\jC\jD}}\) and \(\Qs_{\overline{\jC\jD}}\). Then
\(\Qs_{\jC}\) is extracted from the ``bosonic'' QQ-relation
\eqref{eq:QQf} :
\(\left(\frac{\Qs_{\overline{\jC}}}{\Qs_{\overline{\jD}}}\right)^+-\left(\frac{\Qs_{\overline{\jC}}}{\Qs_{\overline{\jD}}}\right)^-
=
\frac{\Qs_{\overline{\emptyset}}\Qs_{\overline{\jC,\jD}}}{\Qs_{\overline{\jD}}^+\Qs_{\overline{\jD}}^-}\),
which is solved in exactly the same manner as the equation for \(\frac
{\Qs_{\jb}}{\Qs_{\ja}}\).

The fact that \(M_{\overline{\jC\jD}}\) is meromorphic is nontrivial,
and is implied by the polynomiality of \(\mathcal{A}= \frac {Q^-}{F}+F
Q^+= Q^+\left(F+\frac 1 {\mbar F}\right)\) entering there. This polynomiality
  is explained in the footnote \ref{ft:polyn}, and it essentially uses
  the fact that \(\frac {Q^+}{Q^-}=\frac
  {Q^{(+)}}{Q^{(-)}}=\frac{\mbar F}{F}\), which is a consequence of
the asymptotic level matching condition \(\prod_{j=1}^{K_4}
  \frac{x_j^+}{x_j^-}=1\). This proves the meromorphicity of  \(M_{\overline{\jC\jD}}\).

Finally, the expressions of \(\Qs_{\jc},\Qs_{\jD},\Qs_{\jA}\) and
\(\Qs_{\jB}\) are obtained simply by the use of the complex conjugation
transformation on \(\Qs\) functions.

\printindex

\end{document}